\documentclass[aps,prb,reprint,superscriptaddress,showpacs,longbibliography]{revtex4-2}

\usepackage{graphicx}
\usepackage{physics2}
\usephysicsmodule{ab}
\usepackage{amsmath,amsfonts}
\usepackage{tikz}
\usepackage{mathtools}
\usepackage[abs]{overpic}
\usepackage{comment}
\usepackage{ragged2e}
\usepackage[justification=Justified,singlelinecheck=false]{caption}
\usepackage{subcaption}
\usepackage{xcolor}
\usepackage{ulem}
\makeatletter
\newcommand\vb{\@ifstar\boldsymbol\mathbf}

\makeatother

\usepackage{hyperref}
\usepackage{orcidlink}
\hypersetup{
	citecolor = blue,
	colorlinks = true,
	urlcolor = blue
}

\begin{document}

\title{
    Ferromagnetic resonance signatures of edge states for distinguishing chiral and helical \texorpdfstring{$p$}{p}-wave superconductors
}

\author{Shun Muto\,\orcidlink{0009-0009-4922-9678}}
\affiliation{Department of Physics, Nagoya University, Nagoya 464-8602, Japan}
\author{Yuya Ominato\,\orcidlink{0000-0003-2212-9449}}
\affiliation{%
    Waseda Institute for Advanced Study, Waseda University, Shinjuku-ku, Tokyo 169-0051, Japan.
}%
\author{Takeo Kato\,\orcidlink{0000-0001-7984-9354}}
\affiliation{Institute for Solid State Physics, The University of Tokyo, Kashiwa 277-8581, Japan}
\author{Mamoru Matsuo\,\orcidlink{0000-0003-1303-7614}}
\affiliation{%
    Kavli Institute for Theoretical Sciences, University of Chinese Academy of Sciences, Beijing 100190, China.
}%
\affiliation{%
    CAS Center for Excellence in Topological Quantum Computation, University of Chinese Academy of Sciences, Beijing 100190, China
}%
\affiliation{%
    Advanced Science Research Center, Japan Atomic Energy Agency, Tokai 319-1195, Japan
}%
\affiliation{%
    RIKEN Center for Emergent Matter Science (CEMS), Wako, Saitama 351-0198, Japan
}%
\author{Ai Yamakage\,\orcidlink{0000-0003-4052-774X}}
\affiliation{Department of Physics, Nagoya University, Nagoya 464-8602, Japan}

\date{\today}

\begin{abstract}
    We theoretically investigate ferromagnetic resonance (FMR) modulation at interfaces between ferromagnetic insulators and two-dimensional chiral and helical $p$-wave superconductors.
    We evaluate the surface dynamical spin susceptibility, including both edge and bulk quasiparticle excitations.
    For momentum-conserving interfacial exchange at zero temperature, the edge contribution to the Gilbert damping exhibits distinct frequency scalings for the excitation energies well above the spectral broadening, $\hbar\omega\gg2\delta$: $\delta/\ab(\hbar\omega)^{2}$ for the chiral state and $1/\hbar\omega$ for the helical state.
    For a clean superconducting boundary, edge-induced Ising spin anisotropy persists even when interfacial exchange disorder randomizes momentum transfer.
    The low-energy edge response is suppressed when the magnetization aligns with the edge-state spin axis.
    These results establish FMR modulation as a probe complementary to existing methods for distinguishing the chiral and helical states and constraining their $d$-vector directions.
\end{abstract}

\maketitle

\section{Introduction}
Spin-triplet superconductivity supports a rich variety of pairing states and magnetic responses through the spin and orbital degrees of freedom of Cooper pairs~\cite{RevModPhys.63.239}.
In $p$-wave superconductors (pSCs), the $d$-vector describes the momentum and spin structure of Cooper pairs, both of which must be determined to identify the pairing state.
Although quasiparticle spectra and the density of states reveal the excitation structure, they do not directly determine the spin structure, making spin-sensitive measurements essential.
NMR~\cite{ishida1998spin,ishida2008nmr,pustogow2019constraints} and neutron scattering~\cite{PhysRevLett.85.5412,PhysRevLett.125.217004} constrain the spin response, while $\mu$SR~\cite{luke1998time} probes time-reversal symmetry breaking.
Josephson experiments have tested time-reversal symmetry~\cite{PhysRevB.100.094530}, and theoretical proposals use current--phase relations to detect the $d$-vector direction~\cite{zdyn-y6mv}.
Other proposals include the spin-Nernst effect~\cite{PhysRevLett.128.097001} and magnetization-driven spin supercurrents in unitary triplet superconductors under total-spin conservation~\cite{mc66-ymtj}.
Nevertheless, the pairing symmetry of candidate materials such as $\mathrm{Sr_{2}RuO_{4}}$ remains debated~\cite{maeno2024thirty}, motivating complementary probes.

A complementary approach is to probe the existence and properties of boundary excitations in topological superconductors.
For example, gapless Andreev bound states produce subgap features in tunneling conductance, with spectral shapes sensitive to the edge-state dispersion and junction conditions~\cite{KASHIWAYA201425}.
Optical conductivity extends this approach to the charge-current response, with predicted signatures of Majorana boundary modes that depend on the pairing state, measurement geometry, and optical selection rules~\cite{PhysRevLett.126.237002,kamra2024optical,PhysRevB.109.214513}.
For $d$-vector identification, a further question is how these boundary excitations respond to spin perturbations.
In superfluid $^{3}$He-B, theory predicts surface-state contributions to static and dynamical spin responses~\mbox{\cite{PhysRevLett.103.235301,PhysRevB.84.144508,JPSJ.78.123603}}.
Related proposals use impurity-spin ESR to detect changes induced by exchange coupling to Majorana edge modes~\cite{PhysRevB.82.180505}.
These studies motivate using boundary spin dynamics and anisotropy to constrain the $d$-vector, with the interpretation tied to the superconducting state and coupling geometry.

Ferromagnetic resonance (FMR) modulation belongs to this family of spin-sensitive probes.
It shares with NMR and ESR the aim of characterizing spin excitations, using the collective precession of an adjacent ferromagnetic insulator (FI) as the probe.
Under microwave driving, interfacial exchange coupling converts the superconducting spin response into a resonance-field shift $\Delta H$ and a Gilbert damping enhancement $\delta\alpha$.
The latter reflects the dissipative dynamical spin susceptibility transverse to the FI magnetization.
The frequency and magnetization-direction dependences provide information about excitation energies and spin anisotropy relevant to identifying the $d$-vector.
Because the coupling acts at the superconducting surface, both edge and bulk excitations can contribute to the measured response.
This mechanism builds on spin-pumping-induced damping in ferromagnet/normal-metal structures~\cite{PhysRevLett.88.117601}.
In superconductors, this sensitivity to spin dynamics is supported by microscopic theories~\cite{PhysRevB.96.024414,PhysRevB.99.144411,PhysRevB.102.144521} and spin-pumping experiments~\cite{PhysRevB.97.224414,PhysRevApplied.10.014029}.

Interpreting FMR as a pairing probe requires accounting for interface effects: Andreev bound states reshape damping in theory~\cite{PhysRevB.107.144504,PhysRevB.102.180502,PhysRevB.103.L100406}, while experiments reveal dependence on thickness, morphology, and crystal orientation~\cite{yao2021giant,PhysRevB.104.144428,t7vq-dl3f}.
Damping has been interpreted through proximity-induced triplets~\cite{jeon2018enhanced,chan2023controlling} or possible triplet pairing in NbRe~\cite{cirillo2023spin}, while theory supports spin injection mediated by equal-spin pairs~\cite{PhysRevB.107.094513}; these results do not by themselves establish intrinsic pairing symmetry.
For bulk pSCs, theory relates FMR anisotropy to the $d$-vector at flat interfaces, but predicts isotropic responses when exchange-coupling disorder dominates~\cite{PhysRevB.106.L161406}.
This limitation motivates seeking complementary information in boundary spin dynamics, whose FMR signatures have been predicted in proximitized single-subband nanowires~\cite{PhysRevLett.130.237002}.
For three-dimensional topological superconductors, theory predicts Ising-anisotropic FMR from Majorana surface states~\cite{PhysRevB.109.L121405}.
Building on these results, we examine how the combined edge and bulk frequency response constrains the $d$-vector in two-dimensional examples under different interfacial exchange conditions.

We show that FMR retains sensitivity to the $d$-vector through edge-induced Ising spin anisotropy, even when exchange-coupling disorder makes the bulk response isotropic.
We demonstrate this using two-dimensional chiral and helical pSCs as simple examples of distinct $d$-vector structures.
We evaluate $\delta\alpha$ for the FI/pSC bilayer in Fig.~\ref{system}, including edge and bulk quasiparticle excitations in the flat- and rough-interface limits.
Here, interface roughness enters through spatial fluctuations of the exchange coupling.
For a flat interface at zero temperature with the FI magnetization transverse to the edge-spin axis, zero-energy Andreev bound states (\mbox{ZEABS}) yield distinct low-energy frequency dependences, proportional to $\ab(\hbar\omega)^{-2}$ and $\ab(\hbar\omega)^{-1}$ for the chiral and helical states, respectively, at frequencies above the spectral broadening.

\section{Model}\label{Model}
\begin{figure}
  \centering
  \includegraphics[width=\linewidth]{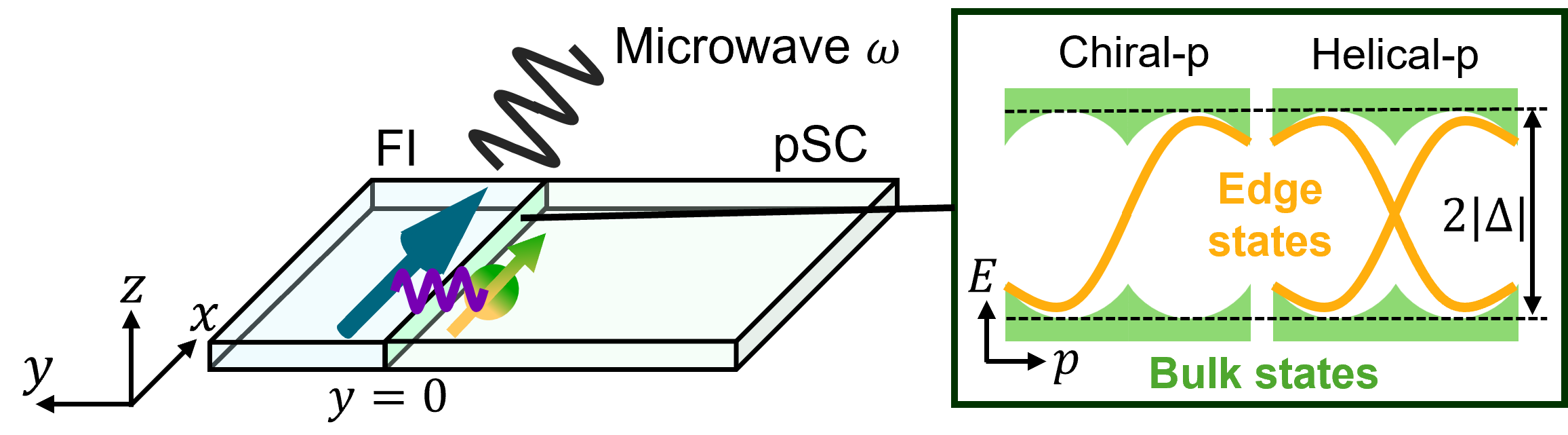}
      \caption{Left: Schematic of the bilayer system consisting of an FI and a pSC. Right: The illustration of the surface spectral functions of chiral-pSC and helical-pSC.}
  \label{system}
\end{figure}

We consider the bilayer system consisting of a ferromagnetic insulator (FI) and a $p$-wave superconductor (pSC) as shown in Fig.~\ref{system}.
For the pSC, we focus on the chiral-pSC and the helical-pSC, whose corresponding $d$-vectors on the square lattice model are given by
\begin{align}
  \vb{d}_{\vb{p}}
  =
  \begin{dcases}
    \ab(0,0,i\sin\ab(p_{x}a)-\sin\ab(p_{y}a)) & :\text{chiral-pSC},
    \\
    \ab(-\sin\ab(p_{y}a),\sin\ab(p_{x}a),0)   & :\text{helical-pSC}.
  \end{dcases}
\end{align}
The FI is interfaced with the $\ab(010)$ surface of the pSC, where the edge states are continuously connected to the bulk electronic structure.
We employ our extended FMR modulation formalism~\cite{qnlf-jx2t}, which can consider the contribution of the bulk and edge states in the pSC on the same footing.
While the edge-state response can be studied within an effective edge Hamiltonian, we aim to examine the FMR response over a broader energy range that encompasses both edge and bulk quasiparticle excitations.
The Hamiltonian is written as
\begin{equation}
  H
  =
  H_{\mathrm{FI}}+H_{p\mathrm{SC}}+H_{\mathrm{int}},
\end{equation}
where $H_{\mathrm{FI}}$, $H_{p\mathrm{SC}}$ and $H_{\mathrm{int}}$ denote the Hamiltonian of an FI, a semi-infinite pSC and interfacial exchange interaction between the FI and the pSC, respectively.
Since the response of FMR modulation is primarily governed by the surface electronic structure of the pSC, we restrict the interfacial coupling to the surface spins of the two materials.
We also assume that each material is clean, such that the in-plane wave vector is conserved on each surface.
In contrast, the interface is disordered due to lattice mismatch, and therefore momentum conservation across the interface is not solely imposed.

\subsection{Ferromagnetic Insulator (FI)}
The first term $H_{\mathrm{FI}}$ is described by the two-dimensional Heisenberg model because the interfacial spin dynamics is dominant.
The Hamiltonian is written as
\begin{align}
  H_{\mathrm{FI}}
   & =
  -\frac{\mathcal{J}}{2}\sum_{\ab\langle  i,j\rangle}\ab(S^{+}_{i}S^{-}_{j}+S^{-}_{i}S^{+}_{j})
  \notag
  \\
   & \qquad-\mathcal{J}\sum_{\ab\langle  i,j\rangle}S^{\nu}_{i}S^{\nu}_{j}-\hbar\gamma  h_{\mathrm{dc}}\sum_{i}S_{i}^{\nu},
\end{align}
where $\mathcal{J}$ denotes the coupling constant between the neighboring localized spins in the FI, $\gamma$ the gyromagnetic ratio, $h_{\mathrm{dc}}$ the static magnetic field, $S^{\nu}_{i}$ the localized spin operator along the magnetization direction $\ab(\nu=x,y,z)$, and $S^{\pm}_{i}$ the corresponding spin ladder operators defined with respect to the same quantization axis, respectively.
Considering the Holstein--Primakoff transformation \cite{PhysRev.58.1098} as
\begin{subequations}
  \begin{align}
    S^{+}_{i}
     & \simeq
    \sqrt{2S_{0}}b_{i},
    \\
    S^{-}_{i}
     & \simeq
    \sqrt{2S_{0}}b^{\dag}_{i},
    \\
    S^{\nu}_{{i}}
     & =
    S_{0}-b^{\dag}_{{i}}b_{{i}},
  \end{align}
\end{subequations}
and the Fourier transformation over the $N_{\mathrm{FI}}$ which means the in-plane sites in the FI:
\begin{align}
  b_{i}
  =\frac{1}{\sqrt{N_{\mathrm{FI}}}}\sum\limits_{\vb{k}}e^{i\vb{k}\cdot\vb{r}_{i}}b_{\vb{k}},
  \quad
  b^{\dag}_{i}
  =\frac{1}{\sqrt{N_{\mathrm{FI}}}}\sum\limits_{\vb{k}}e^{-i\vb{k}\cdot\vb{r}_{i}}b^{\dag}_{\vb{k}},
\end{align}
we derive the effective Hamiltonian within the long-wavelength approximation:
\begin{equation}
  H_{\mathrm{FI}}
  \approx
  \sum\limits_{\vb{k}}\hbar\omega_{\vb{k}}b^{\dag}_{\vb{k}}b_{\vb{k}},
\end{equation}
where $\hbar\omega_{\vb{k}}\approx\mathcal{D}\vb{k}^{2}+\hbar\gamma  h_{\mathrm{dc}}$ is the magnon dispersion with the spin stiffness constant $\mathcal{D}$.

\subsection{\texorpdfstring{$p$}{p}-wave superconductor (pSC)}
The second term $H_{p\mathrm{SC}}$ is described by a semi-infinite tight-binding model along the $-y$ direction on a $\ab(010)$ surface with in-plane momentum $q$ and layer $n$,
\begin{align}
  H_{p\mathrm{SC}}
   & =
  \frac{1}{2}\sum_{q}\sum_{n=0}^{-\infty}\vb{\tilde{c}}^{\dag}_{q,n}\vb{u}\vb{\tilde{c}}_{q,n}
  \notag
  \\
   & \qquad+\frac{1}{2}\sum_{q}\sum_{n=-1}^{-\infty}\ab(\vb{\tilde{c}}^{\dag}_{q,n}\vb{t}\vb{\tilde{c}}_{q,n+1}
  +\mathrm{H.c.}).
  \label{pSCHamiltonian}
\end{align}
$\vb{u}$ and $\vb{t}$ are defined as the onsite and hopping BdG Hamiltonian given by
\begin{equation}
  \vb{u}
  =
  \begin{pmatrix}
    \xi_{q}\sigma^{0} & \Delta_{q}M
    \\
    -\Delta_{q}M      & -\xi_{q}\sigma^{0}
  \end{pmatrix}
  ,\quad
  \vb{t}
  =
  \begin{pmatrix}
    -t\sigma^{0}                            & \displaystyle\frac{i\Delta}{2\sqrt{2}}N
    \\
    \displaystyle\frac{i\Delta}{2\sqrt{2}}N & t\sigma^{0}
  \end{pmatrix},
\end{equation}
\begin{equation}
  \ab(M, N)=
  \begin{cases}
    \ab(\sigma^{x}, \sigma^{x})  & :\text{chiral-pSC},
    \\
    \ab(\sigma^{0}, -\sigma^{z}) & :\text{helical-pSC},
  \end{cases}
\end{equation}
\begin{equation}
  \xi_{q}
  =-2t\cos\ab(qa)-\mu,\quad
  \Delta_{q}
  =\frac{i\Delta}{\sqrt{2}}\sin\ab(qa),
\end{equation}
where $a$ denotes the lattice constant, $t$ nearest-neighbor hopping, $\mu$ chemical potential, and $\Delta$ superconducting gap written as
\begin{equation}
  \Delta=\Delta_{0}\tanh{\ab(1.74\sqrt{\frac{T_{\mathrm{c}}}{T}-1})}.
\end{equation}
$\vb{\tilde{c}}_{q,n}$ $(\vb{\tilde{c}}_{q,n}^\dagger)$ is a quasi-particle annihilation (creation) operator with $q$ at the $n$th layer, which is described in Nambu format as
\begin{subequations}
  \begin{align}
    \vb{\tilde{c}}_{q,n}
     & =
    \,^{t}\ab(c_{q\uparrow,n},c_{q\downarrow,n},c^{\dag}_{-q\uparrow,n},c^{\dag}_{-q\downarrow,n}),
    \\
    \vb{\tilde{c}}^{\dag}_{q,n}
     & =
    \ab(c^{\dag}_{q\uparrow,n},c^{\dag}_{q\downarrow,n},c_{-q\uparrow,n},c_{-q\downarrow,n}).
  \end{align}
\end{subequations}

\subsection{Interfacial exchange coupling}
The third term $H_{\mathrm{int}}$ is the spin transfer Hamiltonian at the interface expressed as
\begin{equation}\label{Hint1}
  H_{\mathrm{int}}
  =
  \sum_{j_{\parallel}}\frac{J_{j_{\parallel}}}{2}\ab(S^{+}_{j_{\parallel}}s^{-}_{j_{\parallel},0}+S^{-}_{j_{\parallel}}s^{+}_{j_{\parallel},0}),
\end{equation}
where $s^{\pm}_{j_{\parallel},0}$ denotes the spin ladder operator at the pSC surface site $\ab(j_{\parallel},0)$, $S^{\pm}_{j_{\parallel}}$ that at the FI site facing it, and $J_{j_{\parallel}}$ the exchange coupling strength at the interface; each surface site of the pSC couples only to the FI site facing it.

Applying the Fourier transformations to the interaction Hamiltonian $H_{\mathrm{int}}$, where $N_{\mathrm{FI}}$ and $N_{\parallel}$ denote the number of in-plane sites in the FI and pSC:
\begin{align}
  S^{\pm}_{j_{\parallel}}
   & =
  \frac{1}{\sqrt{N_{\mathrm{FI}}}}\sum_{\vb k}S_{\vb k}^\pm e^{\pm  i\vb{k}\cdot\vb{r}_{j_{\parallel}}},
  \\
  s^{\pm}_{j_\parallel,0}
   & =
  \frac{1}{N_{\parallel}}\sum_{q}s_{q,0}^\pm e^{\pm  iq\cdot   r_{j_\parallel}},
  \\
  J_{\vb{k},q}
   & =
  \frac{1}{N_{\parallel}\sqrt{N_{\mathrm{FI}}}}\sum\limits_{j_{\parallel}}J_{j_{\parallel}}e^{i\ab(\vb{k}\cdot\vb{r}_{j_{\parallel}}-q\cdot r_{j_{\parallel}})},
\end{align}
we obtain
\begin{equation}
  H_{\mathrm{int}}
  =
  \sum\limits_{\vb{k},q}\ab(\frac{J_{\vb{k},q}}{2}S^{+}_{\vb{k}}s^{-}_{q,0}+\mathrm{H.c.})
\end{equation}

\section{FORMULATION}

\subsection{FMR modulation}
Under microwave driving of the uniform magnon mode: $\vb{k}=\vb{0}$, we define the magnon Green's function in the bilayer system as \cite{PhysRevB.105.205406, PhysRevB.106.L161406, Ominato_2025, qnlf-jx2t}
\begin{align}\label{FIgreen}
   &
  G^{R}_{\vb{k}=\vb{0}}\ab(\omega)
  \notag
  \\
   & \equiv
  -\int^{\beta}_{0}d\tau e^{i \nu_m \tau}\ab\langle    S_{\vb{k}=\vb{0}}^{+}\ab(\tau)S_{\vb{k}=\vb{0}}^{-}(0)\rangle\biggr|_{i\nu_{m}\to \hbar\omega+i0}
  \notag
  \\
   & =
  \frac{2S_{0}/\hbar}{\omega-\omega_{\vb{k}=\vb{0}}+i\alpha\omega-\ab(2S_{0}/\hbar)\Sigma^{R}_{\vb{k}=\vb{0}}\ab(\omega)}.
\end{align}
$\omega_{\vb{k}=\vb{0}}$ and $\alpha$ denote the resonance frequency and the Gilbert damping constant introduced phenomenologically \cite{PhysRevLett.6.223, CHEREPANOV199381, 10.1063/1.5085922}, respectively.
The Gilbert damping constant $\alpha$ corresponds to the linewidth of the absorption spectrum of the FI.
$\Sigma^{R}_{\vb{k}=\vb{0}}\ab(\omega)$ is the magnon self-energy, which is given within the second-order perturbation theory in $H_{\mathrm{int}}$ \cite{PhysRevB.105.205406, PhysRevB.106.L161406, Ominato_2025, qnlf-jx2t} as
\begin{equation}
  \Sigma^{R}_{\vb{k}=\vb{0}}\ab(\omega)
  =
  -\sum\limits_{q}\frac{|J_{\vb{k}=\vb{0},q}|^{2}}{4}\chi^{R}_{q}\ab(\omega),
\end{equation}
where $\chi^{R}_{q}\ab(\omega)$ is the surface dynamic spin susceptibility of the semi-infinite system defined as
\begin{equation}\label{eq:chi_q}
  \chi^{R}_{q}\ab(\omega)
  =\int^{\beta}_{0}d\tau    e^{i\nu_{m}\tau}\ab\langle  s^{+}_{q,0}\ab(\tau)s^{-}_{q,0}\ab(0)\rangle\biggr|_{i  \nu_{m}\to \hbar\omega+i0}.
\end{equation}
Taking into account the effect of interfacial roughness, we write $J_{j_{\parallel}}=J_{1}+\delta J_{j_{\parallel}}$, where $J_{1}$ is the mean coupling and $\delta J_{j_{\parallel}}$ its site-dependent fluctuation, and take the configuration average over $\delta J_{j_{\parallel}}$ with~\cite{PhysRevB.105.205406, PhysRevB.106.L161406, Ominato_2025, qnlf-jx2t}:
\begin{equation}
  \overline{\delta  J_{j_{\parallel}}}
  =
  0,\quad
  \overline{\delta  J_{j_{\parallel}}\delta J_{j^{\prime}_{\parallel}}}
  =
  J^{2}_{2}\delta_{j_{\parallel},j^{\prime}_{\parallel}}.
\end{equation}
In this paper, we focus on the enhanced Gilbert damping constant $\delta\alpha\ab(\omega)$, which is expressed as
\begin{align}\label{damping}
  \delta\alpha\ab(\omega)
   & \equiv
  -\frac{2S_{0}}{\hbar\omega}\mathrm{Im}\overline{\Sigma^{R}_{\vb{k}=\vb{0}}\ab(\omega)}
  \notag
  \\
   & =
  \frac{2S_{0}}{\hbar\omega}\ab(\frac{J^{2}_{1}}{4N_{\mathrm{FI}}}\mathrm{Im}\chi_{\mathrm{uni}}\ab(\omega)+\frac{J^{2}_{2}}{4N_{\mathrm{FI}}}\mathrm{Im}\chi_{\mathrm{loc}}\ab(\omega)),
\end{align}
where $\chi_{\mathrm{uni}}\ab(\omega)$ and $\chi_{\mathrm{loc}}\ab(\omega)$ denote the uniform and local spin susceptibility given by
\begin{equation}
  \chi_{\mathrm{uni}}\ab(\omega)
  =
  \chi^{R}_{q=0}\ab(\omega)
  ,\quad
  \chi_{\mathrm{loc}}\ab(\omega)
  =
  \frac{1}{N_{\parallel}}\sum_{q}\chi^{R}_{q}\ab(\omega).
\end{equation}
Accordingly, we denote the first and second terms on the right-hand side of Eq.~\eqref{damping} by $\delta\alpha_{\mathrm{uni}}\ab(\omega)$ and $\delta\alpha_{\mathrm{loc}}\ab(\omega)$, respectively.

\subsection{Surface spin susceptibility}
We evaluate the quasiparticle spin response expressed in Eq.~\eqref{eq:chi_q} with the superconducting order parameter held fixed.
Collective order-parameter dynamics, including the surface Cooper-pair spin waves predicted for triplet superconductors~\cite{PhysRevLett.129.237002}, are beyond the scope of the present calculation.
Hybridization between finite-wave-vector magnons and superconducting phase modes has also been predicted for a ferromagnetic insulator coupled to a two-dimensional $s$-wave superconducting surface with spin--momentum locking~\cite{PhysRevB.113.104514}.

The spin ladder operators $s^{\pm}_{q,0}$, corresponding to the FI magnetization direction $\nu$, are defined as
\begin{equation}
  s^{\pm}_{q,0}
  =\frac{1}{2}\sum\limits_{p}\sum_{ss^{\prime}}c^{\dag}_{ps,0}\ab(\sigma^{\pm})_{ss^{\prime}}c_{p\pm   qs^{\prime},0},
\end{equation}
where $s,s^{\prime}$ denote the spin indices $\uparrow,\downarrow$ and $\sigma^{\pm}$ the $2\times2$ matrices expressed with the Pauli matrices,
\begin{align}\label{sigmapm}
  \sigma^{\pm}
  =
  \begin{dcases}
    \sigma^{x}\pm    i\sigma^{y}:\,\nu=z,
    \\
    \sigma^{y}\pm    i\sigma^{z}:\,\nu=x,
    \\
    \sigma^{z}\pm    i\sigma^{x}:\,\nu=y.
  \end{dcases}
\end{align}
Substituting Eq.~\eqref{sigmapm} into Eq.~\eqref{eq:chi_q}, we derive the imaginary part of the surface spin susceptibility $\chi^{R}_{q}\ab(\omega)$ with $\nu=x,y,z$ using the surface Gor'kov--Green function $\tilde{g}$.
The Green's function is defined with the normal Green's function $g$ and the anomalous Green's function $f$ as
\begin{align}\label{GGF}
  \tilde{g}^{R}_{p}\ab(E)
   & =
  \int^{\beta}_{0}d\tau
  e^{i {\epsilon_{n}} \tau}\ab[-\ab\langle    \vb{\tilde{c}}_{p,0}\ab(\tau)\vb{\tilde{c}}^{\dag}_{p,0}\ab(0)\rangle]\biggr|_{i {\epsilon_n} \to {E}+i0}
  \notag
  \\
   & =
  \begin{pmatrix}
    g^{R}_{p}\ab(E)       & f^{R}_{p}\ab(E)
    \\[1ex]
    \bar{f}^{R}_{p}\ab(E) & \bar{g}^{R}_{p}\ab(E)
  \end{pmatrix}.
\end{align}
We apply the recursive method~\cite{umerski1997closed} for the derivation of the surface Green's function and the details are given in Appendix~\ref{SGF}.
For both superconductors, the surface spin susceptibility can be written in the common form
\begin{align}\label{eq:surface-chi-common}
   & \mathrm{Im}\chi^{R,\eta}_{q,\nu}\ab(\omega)
  \notag                                                                                   \\
   & =\sum_{p}\int\frac{dE}{\pi}\ab[n_{\mathrm{F}}\ab(E)-n_{\mathrm{F}}\ab(E+\hbar\omega)]
  \mathcal{K}^{\eta}_{\nu}\ab(p,E;q,\omega),
\end{align}
where $\eta=\mathrm{c},\mathrm{h}$ denotes the chiral-pSC and helical-pSC, respectively, and $\nu$ is the FI magnetization direction.
For each superconducting state, we define $A_s=\mathrm{Im}[g^R_{pss}(E)]$ and $B_{ss'}=\mathrm{Re}[f^R_{pss'}(E)]$, with primed quantities evaluated at $(p+q,E+\hbar\omega)$.
The kernels are
\begin{equation}\label{eq:surface-chi-kernels}
  \begin{aligned}
    \mathcal{K}^{\mathrm{c}}_{\nu}
     & =A_{\uparrow}A'_{\uparrow}
    -\delta_{\nu z}B_{\uparrow\downarrow}B'_{\uparrow\downarrow},
    \\[0.5ex]
    \mathcal{K}^{\mathrm{h}}_{\nu}
     & =\begin{cases}
          \begin{aligned}
        \frac{1}{4}\Big[ & (A_{\uparrow}+A_{\downarrow})(A'_{\uparrow}+A'_{\downarrow}) \\
                         & -(B_{\uparrow\uparrow}\mp B_{\downarrow\downarrow})
          (B'_{\uparrow\uparrow}\mp B'_{\downarrow\downarrow})\Big],
      \end{aligned}
                                       & \nu=x,y, \\[0.5ex]
          A_{\uparrow}A'_{\downarrow}, & \nu=z.
        \end{cases}
  \end{aligned}
\end{equation}
Here, $\delta_{\nu z}$ is the Kronecker delta, and the upper (lower) sign corresponds to $\nu=x$ ($\nu=y$).
The detailed derivations are given in Appendix~\ref{pcal}.
\section{Results}\label{Results}
We calculate the Gilbert damping enhancement $\delta\alpha\ab(\omega)$ using the parameter set $a=1$, $\Delta_{0}=1.76k_{\mathrm{B}}T_{\mathrm{c}}$, $t/\Delta_{0}=\mu/\Delta_{0}=100$, and $\delta/\Delta_{0}=0.001$.
We compute $\delta\alpha_{\mathrm{uni}}\ab(\omega)$ and $\delta\alpha_{\mathrm{loc}}\ab(\omega)$ separately in the following subsections and present results as $\delta\alpha_{\mathrm{uni}}/\delta\alpha_{\mathrm{uni,\,N}}$ and $\delta\alpha_{\mathrm{loc}}/\delta\alpha_{\mathrm{loc,\,N}}$, respectively.
Here, $\delta\alpha_{\mathrm{uni,\,N}}\equiv\lim_{\omega\to0}\delta\alpha_{\mathrm{uni}}\ab(\omega,T=T_{\mathrm{c}})$ and $\delta\alpha_{\mathrm{loc,\,N}}\equiv\lim_{\omega\to0}\delta\alpha_{\mathrm{loc}}\ab(\omega,T=T_{\mathrm{c}})$ denote the corresponding normal-state values in the static limit.
The total damping enhancement, including both the uniform and local contributions, is discussed in Sec.~\ref{discussion}.

\begin{figure}
  \begin{minipage}{0.49\linewidth}
    \subcaption{Chiral}
    \includegraphics[width=\columnwidth]{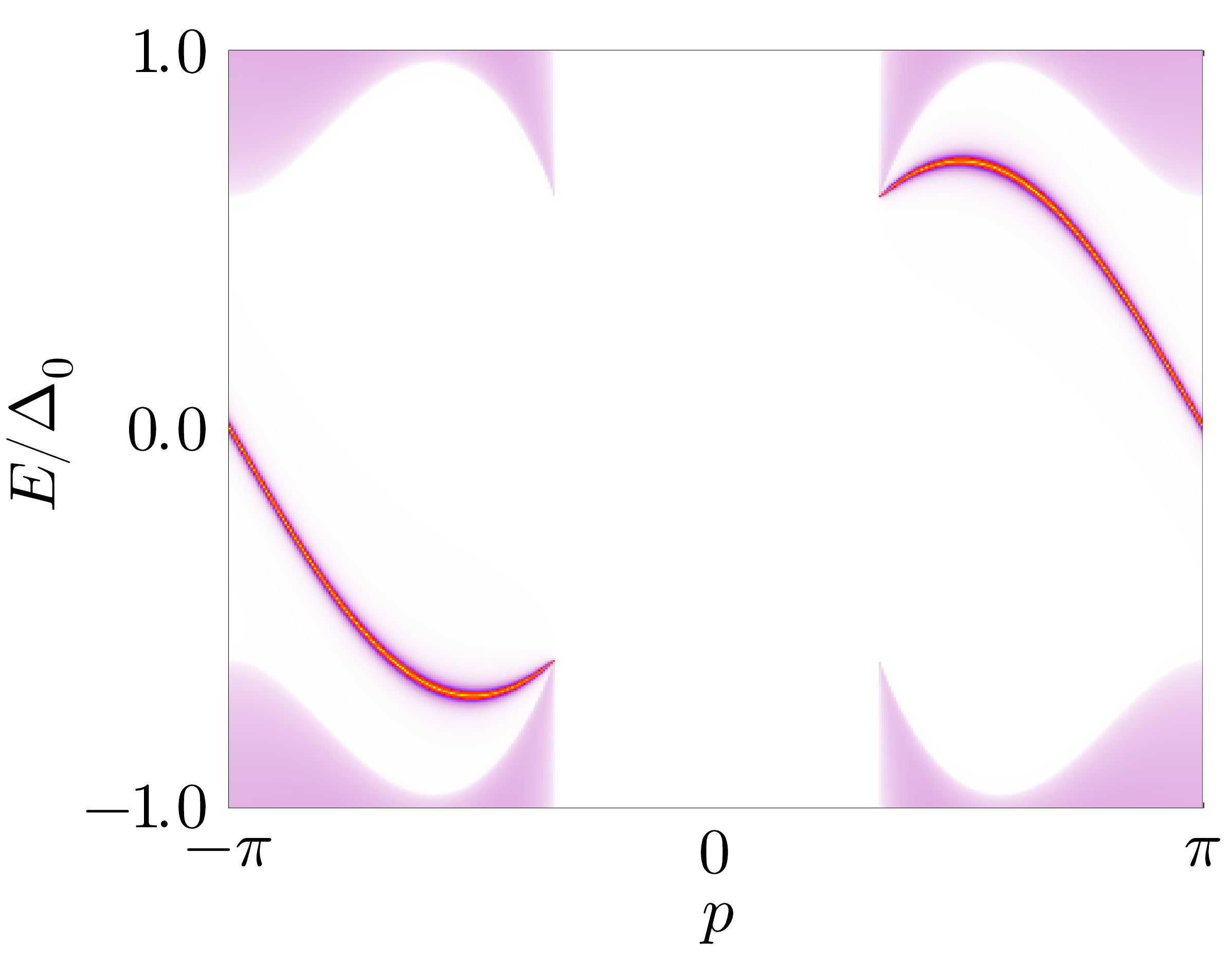}
    \label{cpspectrum}
  \end{minipage}
  \begin{minipage}{0.49\linewidth}
    \subcaption{Helical}
    \includegraphics[width=\columnwidth]{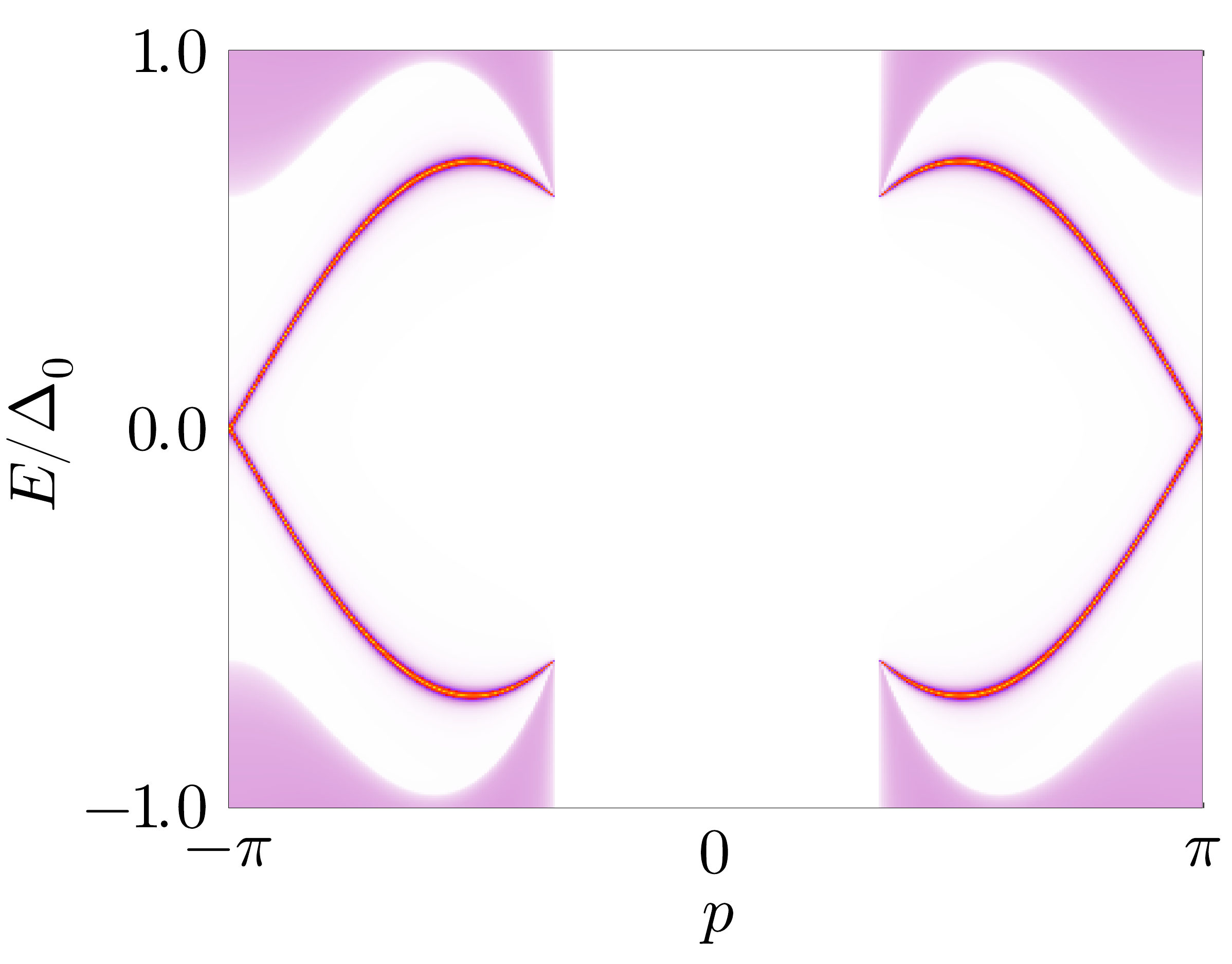}
    \label{hpspectrum}
  \end{minipage}
  \caption{The surface spectra of the chiral-pSC and helical-pSC, respectively. These edge bands exhibit nonlinear dispersion within the bulk gap.}
\end{figure}

To characterize the bulk and edge states relevant to $\delta\alpha\ab(\omega)$, we also calculate the spectral functions $\rho\ab(E)$ defined as
\begin{equation}
  \rho\ab(E)
  =
  -\frac{1}{\pi}\mathrm{Im}\ab[g^{R}_{p\uparrow\uparrow}\ab(E)+g^{R}_{p\downarrow\downarrow}\ab(E)].
\end{equation}
The surface spectral functions for the same parameter set are shown in Figs.~\ref{cpspectrum} and \ref{hpspectrum} for the chiral and helical states, respectively.
Both exhibit edge bands within the bulk gap.
For the present lattice model, the edge dispersions are  $E^{\mathrm{c}}_{p\uparrow}=E^{\mathrm{c}}_{p\downarrow}=D_p$ for the chiral state and $E^{\mathrm{h}}_{p\uparrow}=-D_p$, $E^{\mathrm{h}}_{p\downarrow}=D_p$ for the helical state, where $D_p=(\Delta/\sqrt{2})\sin(pa)$.
For the parameters used here, the edge bound states exist for $|p|>\pi/(3a)$ and merge into the bulk continuum at $|p|=\pi/(3a)$.
For $|p|<\pi/3a$, the normal-state band does not cross the Fermi level, so the bulk continuum lies far outside the energy range shown; for the parameters used here, $\mu/\Delta_{0}=100$, yielding a minimum bulk quasiparticle energy of $E=\mu=100\Delta_{0}$ at $p=0$.

\subsection{Uniform spin susceptibility}\label{uss}
\begin{figure}
  \begin{subfigure}{\columnwidth}
    \caption{Chiral}
    \includegraphics[width=0.8\columnwidth]{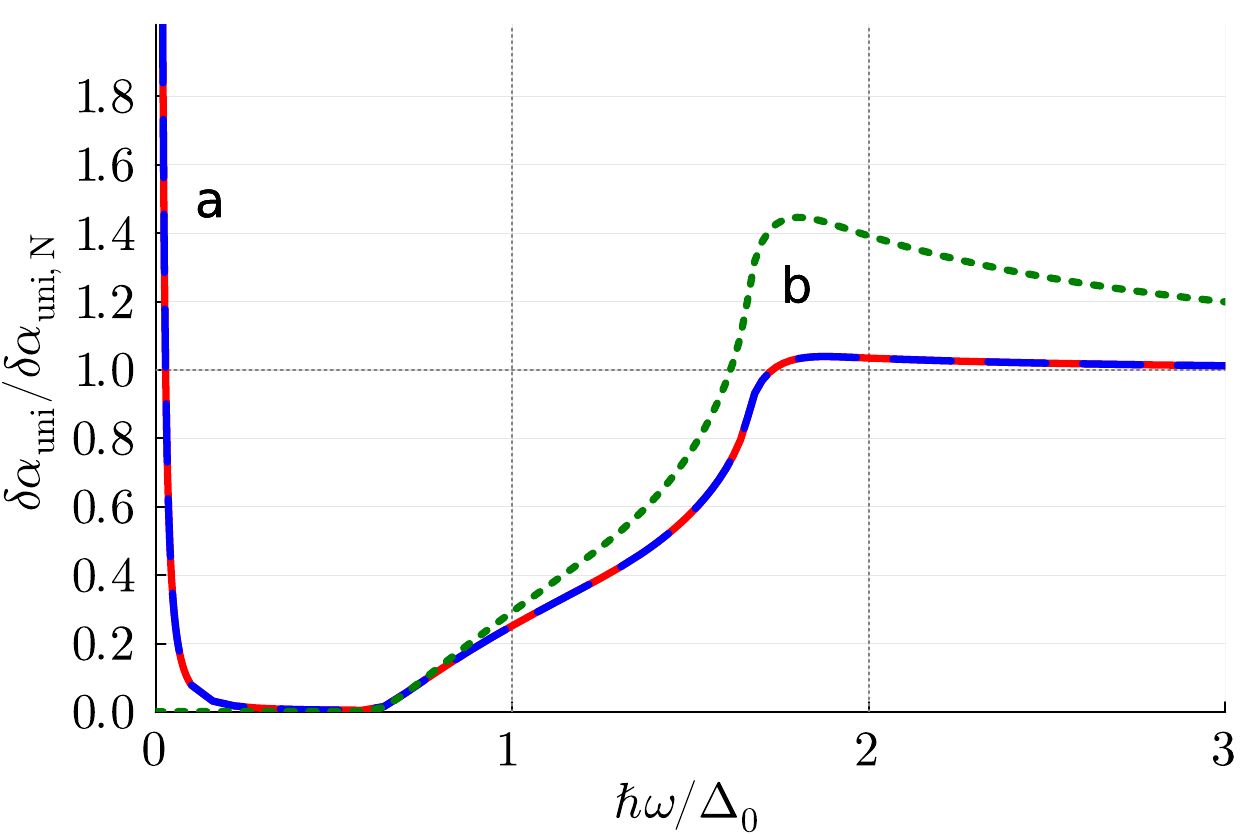}
    \label{cpGDCuni}
  \end{subfigure}

  \begin{subfigure}{\columnwidth}
    \caption{Helical}
    \includegraphics[width=0.8\columnwidth]{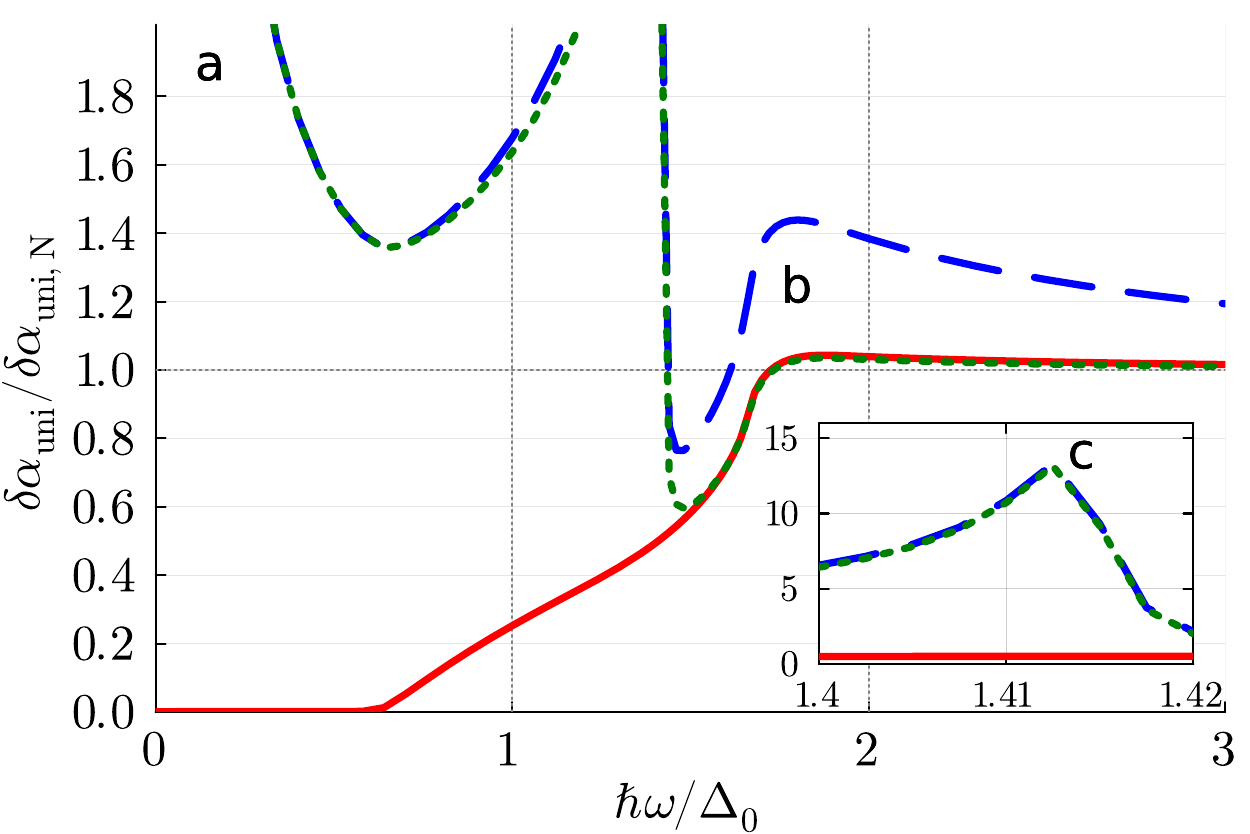}
    \label{hpGDCuni}
  \end{subfigure}
  \par\vspace{1.0em}
  \makebox[\linewidth]{
    \includegraphics[trim=12cm 0.3cm 12cm 0.3cm,clip,scale=0.45]{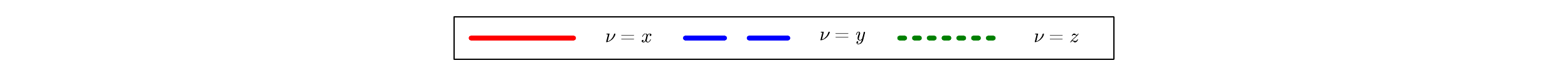}}
  \vspace{0.5em}

  \begin{subfigure}{\columnwidth}
    \caption{}
    \includegraphics[width=0.8\columnwidth]{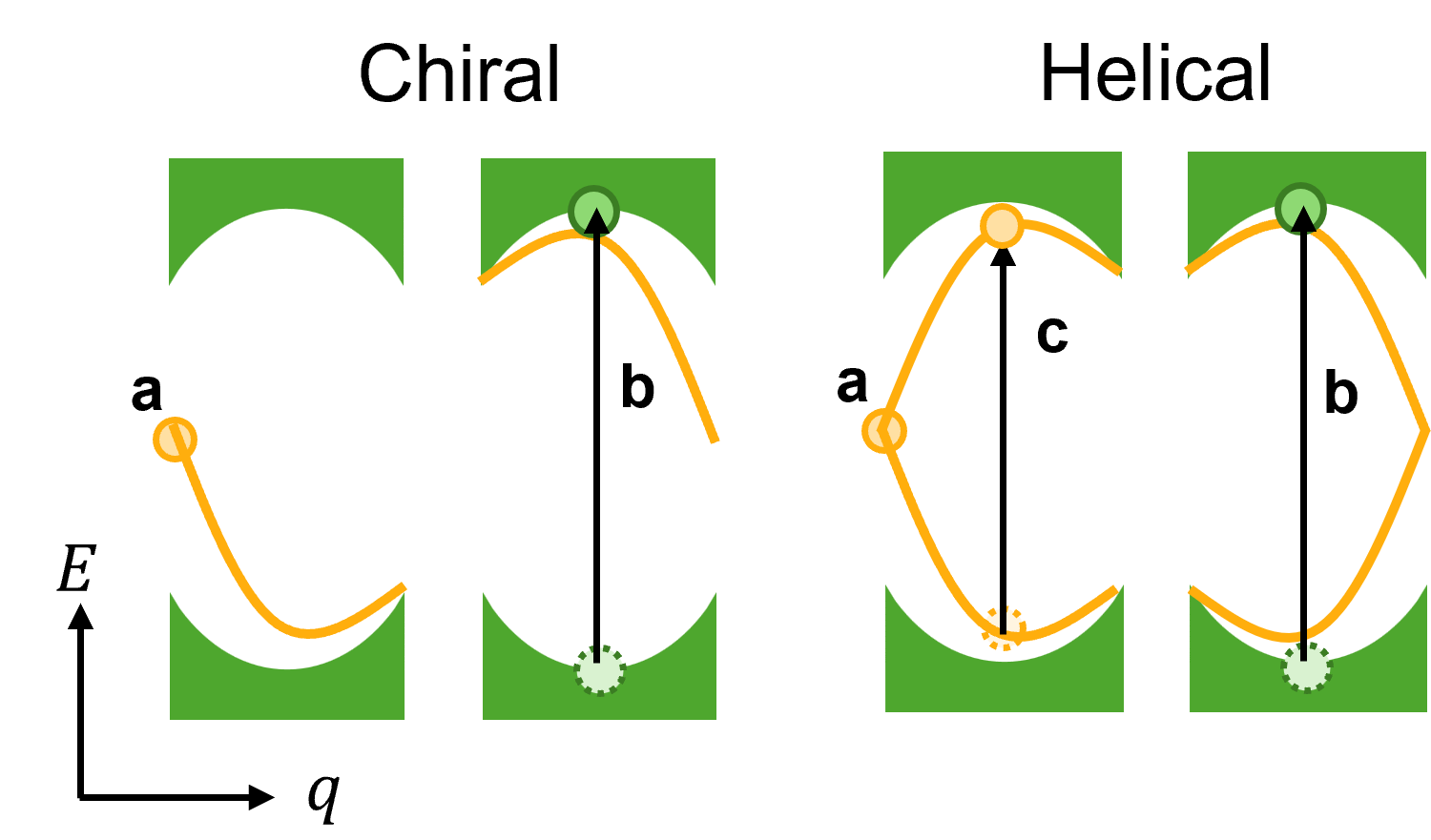}
    \label{p+ipexcitationuni}
  \end{subfigure}
  \caption{Frequency dependence of $\delta\alpha_{\mathrm{uni}}\ab(\omega)$ for $T=0$ along $\nu=x,y,z$ of FI magnetization for the chiral-pSC and helical-pSC in Figs.~\ref{cpGDCuni} and \ref{hpGDCuni}, respectively. The label ``a", ``b" and ``c" correspond to the schematic illustration of each excitation in Fig.~\ref{p+ipexcitationuni}.}
  \label{GDCuni}
\end{figure}

We calculate the frequency dependence of the uniform spin susceptibility $\chi_{\mathrm{uni}}\ab(\omega)$ by setting $q=0$ in Eq.~\eqref{eq:surface-chi-common} with the kernels in Eq.~\eqref{eq:surface-chi-kernels}.
This quantity probes excitations between quasiparticles with the same wave vector.
The frequency dependence of $\delta\alpha_{\mathrm{uni}}\ab(\omega)$ at $T=0$ for the chiral-pSC and helical-pSC is shown in Figs.~\ref{cpGDCuni} and \ref{hpGDCuni}.
Both the chiral-pSC with $\nu=x,y$ and the helical-pSC with $\nu=y, z$ exhibit a peak near zero frequency originating from transitions between the zero-energy Andreev bound states (ZEABS), labeled ``a'' on the left and right sides of Fig.~\ref{p+ipexcitationuni}, respectively.
Their frequency and temperature dependences can be discussed in terms of the effective edge spectra in the low-energy regime
\begin{align}\label{eq:edge-spectral-common}
  -\mathrm{Im}\ab[g^{R,\eta}_{pss}\ab(E)]
                       & \sim L_{\delta}\ab(E-E^{\eta}_{ps}),
  \notag                                                      \\
  L_{\delta}\ab(x)     & =\frac{\delta}{x^{2}+\delta^{2}},
  \notag                                                      \\
  E^{\eta}_{p\uparrow} & =\zeta_{\eta}vp,
  \qquad E^{\eta}_{p\downarrow}=vp,
\end{align}
where $\zeta_{\mathrm{c}}=+1$ and $\zeta_{\mathrm{h}}=-1$ describe copropagating and counterpropagating branches, respectively, and $p$ is measured from the zero-energy crossing.
We hold the edge velocity, momentum-independent spectral weights, and broadening fixed to isolate the effects of frequency and Fermi occupation from changes in the spectra caused by $\Delta(T)$.
For $q=0$ and the magnetization directions specified above, Eq.~\eqref{eq:surface-chi-common} gives the edge contribution at finite temperature as
\begin{align}\label{eq:uniform-edge-scaling}
   & \delta\alpha^{\eta}_{\mathrm{uni,edge}}\ab(\omega,T)
  \notag                                                     \\
   & \propto\frac{1}{\hbar\omega}\int_{-\infty}^{\infty}dE\,
  \ab[n_{\mathrm{F}}\ab(E)-n_{\mathrm{F}}\ab(E+\hbar\omega)]
  \mathcal{I}_{\eta}\ab(E,\hbar\omega).
\end{align}
Here we take $\omega>0$; the proportionality constants include the fixed spectral weights and the normalization of the momentum sum.
\begin{align}\label{eq:uniform-edge-overlap}
  \mathcal{I}_{\eta}\ab(E,\hbar\omega)
   & \equiv\frac{1}{\pi}\sum_{p}\,
  \mathcal{K}^{\eta}_{\nu}\ab(p,E;q=0,\omega)
  \notag                           \\
   & 
   \propto
   \frac{1}{|v|}
  \begin{cases}
    L_{2\delta}\ab(\hbar\omega),    & \eta=\mathrm{c}, \\[0.5ex]
    L_{2\delta}\ab(2E+\hbar\omega), & \eta=\mathrm{h}.
  \end{cases}
\end{align}

For the chiral state, the momentum-integrated edge kernel is independent of energy.
Using $\int_{-\infty}^{\infty}dE\,\ab[n_{\mathrm{F}}\ab(E)-n_{\mathrm{F}}\ab(E+\hbar\omega)]=\hbar\omega$, we obtain
\begin{equation}\label{eq:uniform-chiral-edge}
  \delta\alpha^{\mathrm{c}}_{\mathrm{uni,edge}}\ab(\omega,T)
  \propto\frac{\delta}{(\hbar\omega)^{2}+4 \delta^{2}}.
\end{equation}
Thus, the occupation factors introduce no temperature dependence, provided that the thermal and frequency windows remain within the range of linear dispersion and approximately constant spectral weights.
For $2\delta\ll\hbar\omega$, the overlap of the Lorentzian tails of the copropagating branches gives the $\delta/(\hbar\omega)^{2}$ dependence.

For the helical state, the kernel instead selects transitions between energies near $-\hbar\omega/2$ and $+\hbar\omega/2$.
In the narrow-width regime $\delta\ll\hbar\omega$, the energy integral gives \begin{equation}\label{eq:uniform-helical-edge}
  \delta\alpha^{\mathrm{h}}_{\mathrm{uni,edge}}\ab(\omega,T)
  \propto\frac{1}{\hbar\omega}\tanh\ab(\frac{\hbar\omega}{4k_{\mathrm{B}}T}).
\end{equation}
At fixed frequency, the response approaches a constant as $T\to0$ and varies as $1/T$ for $2\delta\ll\hbar\omega\ll 4k_{\mathrm{B}}T$, with the edge spectral parameters held fixed.
At $T=0$, the energy integral can also be evaluated without assuming $\delta\ll\hbar\omega$.
Within the same linear-edge approximation, we obtain
\begin{equation}\label{eq:uniform-helical-edge-finite-width}
    \delta\alpha^{\mathrm{h}}_{\mathrm{uni,edge}}\ab(\omega,T=0)
    \propto
    \frac{1}{\hbar\omega}\arctan\ab(\frac{\hbar\omega}{2\delta}).
\end{equation}
This expression recovers the $1/(\hbar\omega)$ dependence for $\hbar\omega\gg2\delta$ and, for fixed finite $\delta$, approaches a frequency-independent limit for $\hbar\omega\ll2\delta$, as does the chiral contribution.

Beyond the low-frequency regime, the chiral edge-to-edge response drops to a value close to zero.
$\delta\alpha_{\mathrm{uni}}\ab(\omega)$ also gradually increases from $\hbar\omega/\Delta_{0}\sim0.6$, due to the excitation between the ZEABS and the bulk band, eventually emerging into the bulk-to-bulk excitation feature around $\hbar\omega/\Delta_{0}\sim1.8$, which is denoted ``b" on the left side of Fig.~\ref{p+ipexcitationuni}.
With $\nu=z$, the edge-to-edge excitation peak is strongly suppressed, whereas the bulk-to-bulk excitation, which is denoted ``b" on the left side of Fig.~\ref{p+ipexcitationuni} is stronger than those with $\nu=x,y$.

For the helical-pSC, edge-to-edge transitions remain allowed within the bulk gap beyond the low-frequency peak.
At higher frequencies, edge-to-bulk excitations gradually enhance $\delta\alpha_{\mathrm{uni}}\ab(\omega)$, while another edge-to-edge peak occurs at $\hbar\omega/\Delta_{0}\sim1.4$, labeled ``c" on the right side of Fig.~\ref{p+ipexcitationuni}.
The bulk-to-bulk feature appears around $\hbar\omega/\Delta_{0}\sim1.8$, labeled ``b" on the right side of Fig.~\ref{p+ipexcitationuni}.
With $\nu=y$, the behavior of $\delta\alpha_{\mathrm{uni}}\ab(\omega)$ is similar to that in the $\nu=z$ case, but the bulk-to-bulk excitation, which is denoted ``b" on the right side of Fig.~\ref{p+ipexcitationuni} is larger than that in the $\nu=z$ case.
With $\nu=x$, the edge-to-edge excitation peak is strongly suppressed, and the bulk-to-bulk excitation, which is denoted ``b" on the right side of Fig.~\ref{p+ipexcitationuni} remains the same value as $\nu=z$.

For the helical state, the magnetization-direction dependence of the bulk-to-bulk excitation differs from the bulk result of Ref.~\cite{PhysRevB.106.L161406}, where the damping is larger for in-plane magnetization ($\nu=x$; $\nu=y$ is equivalent in the bulk by the fourfold symmetry) than for $\nu=z$.
Here the enhancement appears only for $\nu=y$, whereas the values for $\nu=x$ and $\nu=z$ coincide.
A possible origin of this difference is the interface geometry.
In Ref.~\cite{PhysRevB.106.L161406} the FI covers the two-dimensional plane and the uniform response conserves the two-dimensional momentum, whereas here the FI couples to the edge, so that only the momentum along the edge is conserved and the surface response involves transitions between all momenta normal to the edge.
A quantitative verification of this effect is left for future work.

\begin{figure}
  \centering

  \begin{minipage}{0.48\linewidth}
    \subcaption{Chiral, $\hbar\omega/\Delta_{0}=0.04$}
    \includegraphics[width=\columnwidth]{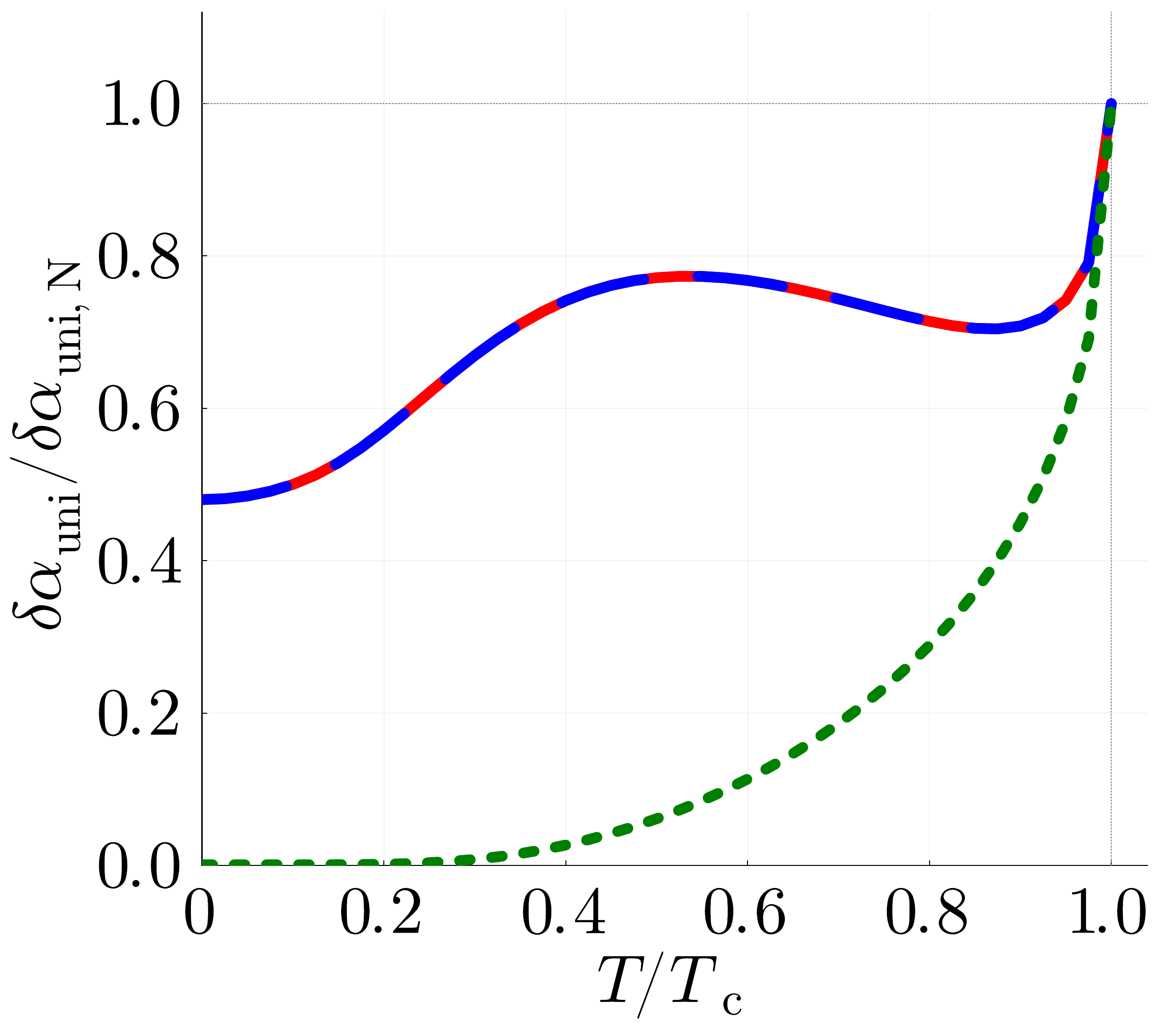}
    \label{cpGDCunitemp0.04}
  \end{minipage}
  \begin{minipage}{0.48\linewidth}
    \subcaption{Chiral, $\hbar\omega/\Delta_{0}=1.3$}
    \includegraphics[width=\columnwidth]{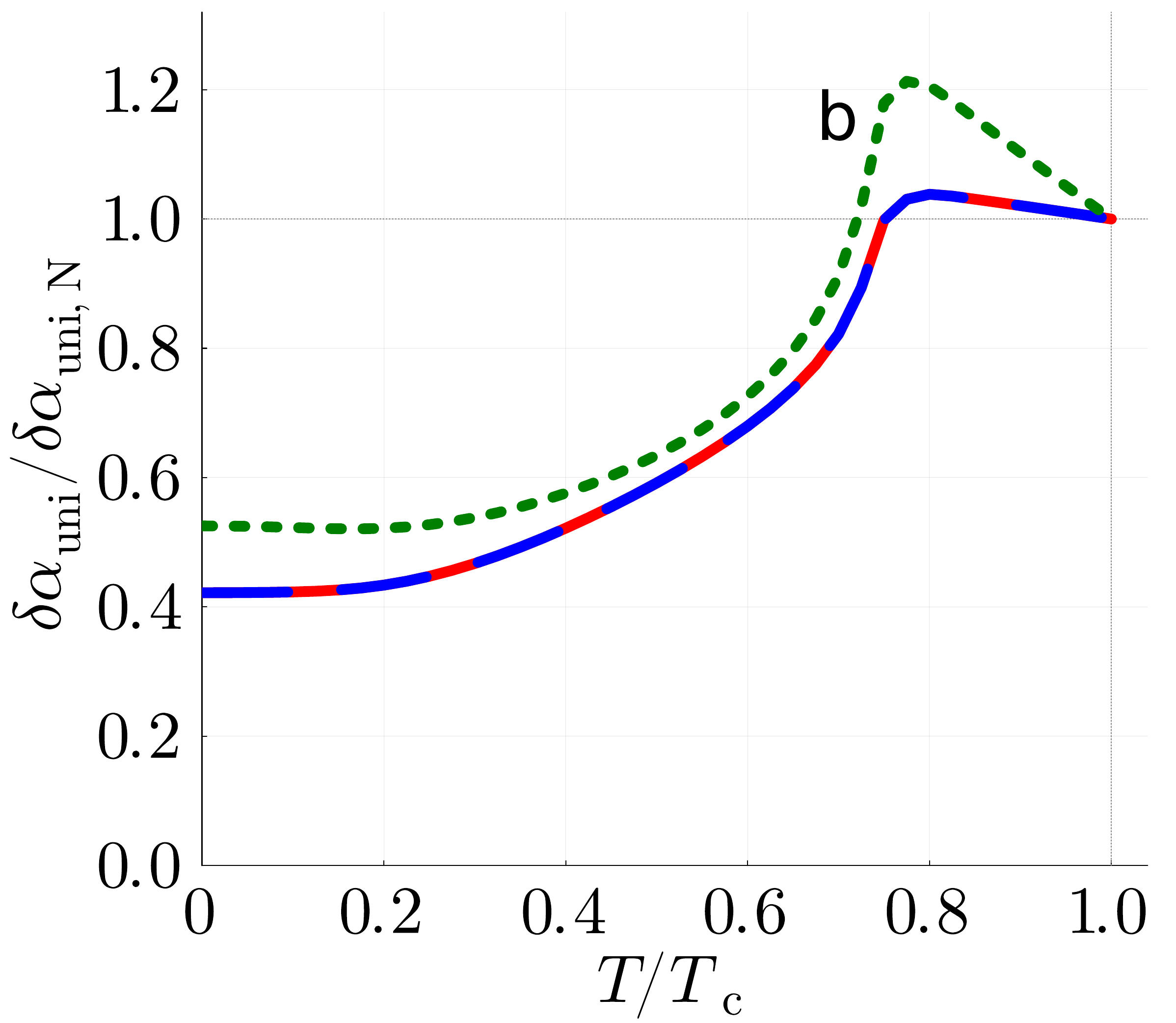}
    \label{cpGDCunitemp1.3}
  \end{minipage}
  \begin{minipage}{0.48\linewidth}
    \subcaption{Helical, $\hbar\omega/\Delta_{0}=0.04$}
    \includegraphics[width=\linewidth]{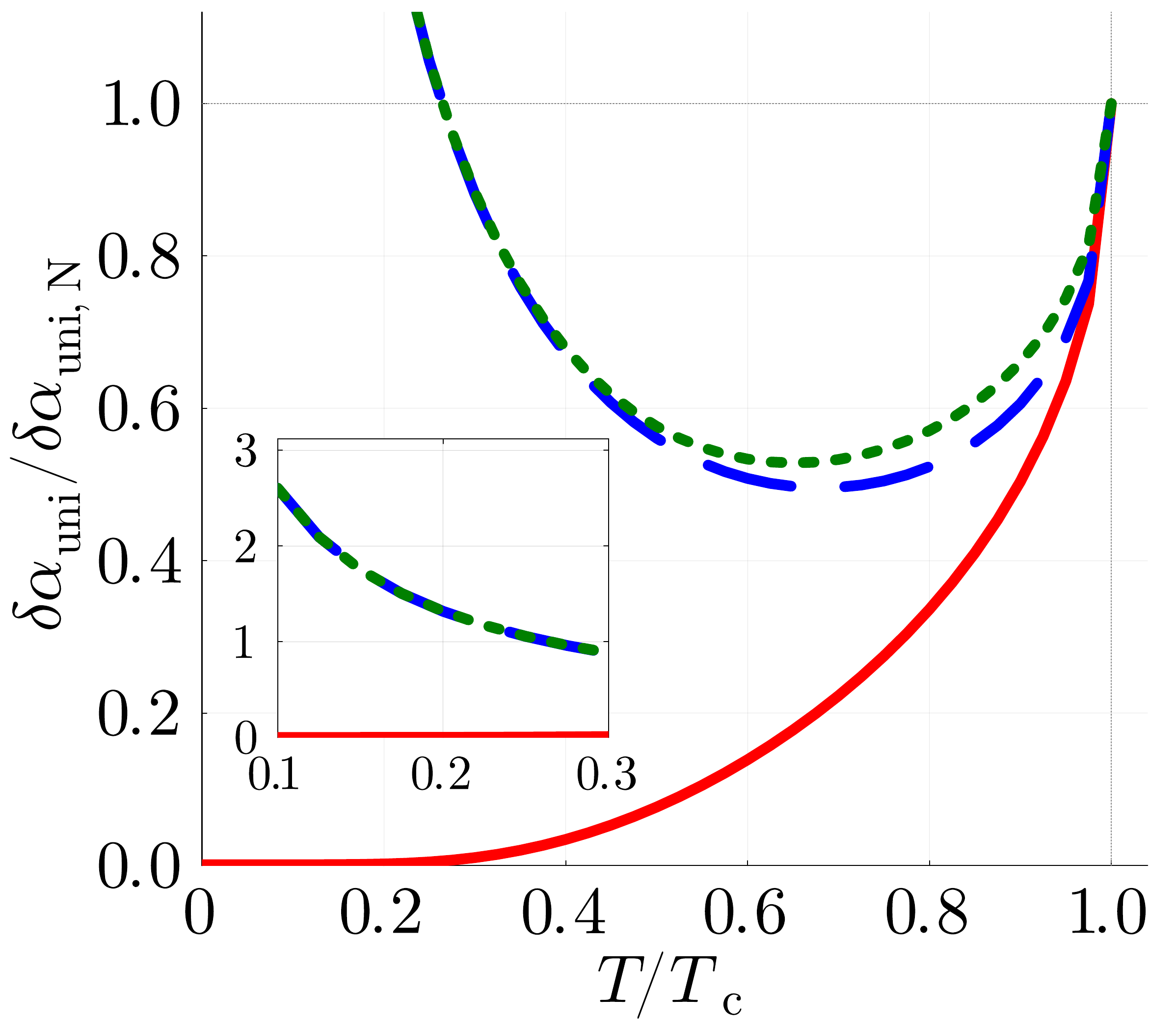}
    \label{hpGDCunitemp0.04}
  \end{minipage}
  \begin{minipage}{0.48\linewidth}
    \subcaption{Helical, $\hbar\omega/\Delta_{0}=1.3$}   \includegraphics[width=\linewidth]{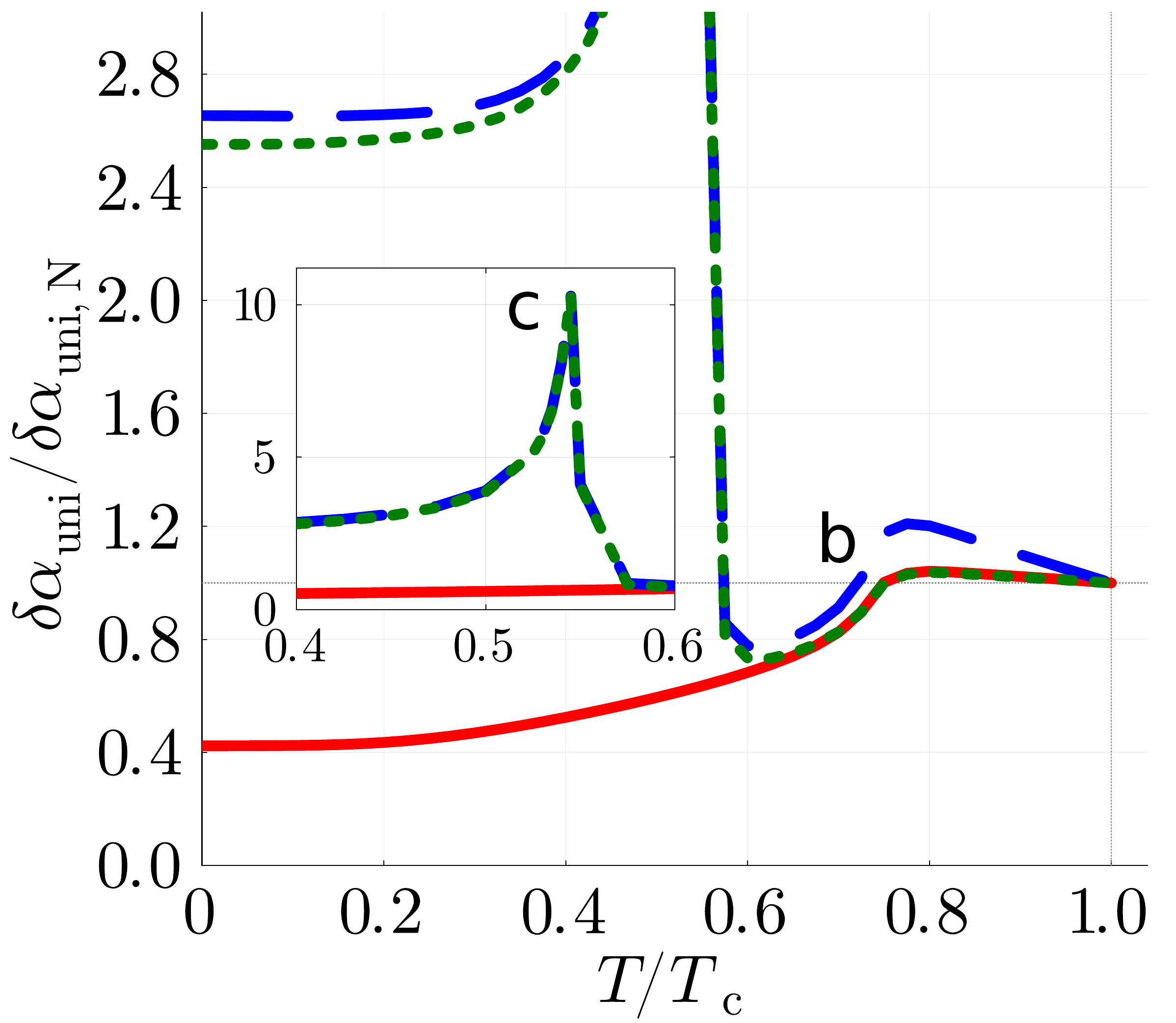}
    \label{hpGDCunitemp1.3}
  \end{minipage}
  \makebox[\linewidth]{%
    \includegraphics[trim=12cm 0.3cm 12cm 0.3cm,clip,scale=0.45]{Figure/20260216pSC,legend,axislabel.pdf}}
  \caption{Temperature dependence of $\delta\alpha_{\mathrm{uni}}\ab(\omega)$ for $\hbar\omega/\Delta_{0}=0.04\ \mathrm{and}\ 1.3$ along $\nu=x,y,z$ of FI
    magnetization for the chiral-pSC and helical-pSC.}
  \label{GDCunitemp}
\end{figure}

The temperature dependence of $\delta\alpha_{\mathrm{uni}}\ab(\omega)$ at low frequency $\ab(\hbar\omega/\Delta_{0}=0.04)$ and intermediate frequency $\ab(\hbar\omega/\Delta_{0}=1.3)$ is shown in Fig.~\ref{GDCunitemp} for chiral-pSC and helical-pSC.
For the low-frequency curves in Figs.~\ref{cpGDCunitemp0.04} and \ref{hpGDCunitemp0.04}, Eqs.~\eqref{eq:uniform-chiral-edge} and \eqref{eq:uniform-helical-edge} describe the occupation-factor dependence of the specified edge contributions at fixed spectral parameters.
The numerical calculation uses a temperature-independent broadening, $\delta/\Delta_0=0.001$, but includes the temperature dependence of the superconducting gap.
Changes in $\Delta(T)$ modify the edge velocities and spectral weights, while transitions involving bulk states also contribute.
The idealized temperature dependences therefore do not by themselves describe the full numerical curves over the entire temperature range.

On the other hand, at the frequency $\hbar\omega/\Delta_{0}=1.3$, shown in Figs.~\ref{cpGDCunitemp1.3} and \ref{hpGDCunitemp1.3},
as the superconducting gap decreases with increasing temperature, the specific frequencies of the corresponding excitations also shift toward lower frequencies, allowing their contributions to be captured in the present frequency range.
For chiral-pSC, the temperature dependence can capture the bulk-to-bulk excitation labeled ``b", whereas for helical-pSC, it captures both the edge bands excitation labeled ``c" and the bulk-to-bulk excitation labeled ``b".

\subsection{Local spin susceptibility}\label{lss}
We next consider the local spin susceptibility $\chi_{\mathrm{loc}}\ab(\omega)$ by summing Eq.~\eqref{eq:surface-chi-common} over $q$ and dividing by $N_{\parallel}$.
In this case, $\chi_{\mathrm{loc}}\ab(\omega)$ can allow the excitation process of not only the same wave number quasiparticle, but also the other wave number quasiparticle.
Summing over the two wave numbers, we obtain
\begin{align}\label{eq:surface-chiloc}
   & \mathrm{Im}\chi_{\mathrm{loc}}\ab(\omega)
  \notag
  \\
   & =
  \int\frac{dE}{\pi}\ab(n_{\mathrm{F}}\ab(E)-n_{\mathrm{F}}\ab(E+\hbar\omega))
  \notag
  \\
   & \qquad\times
  \begin{dcases}
    X_{\uparrow}X^{\prime}_{\uparrow}-\delta_{\nu  z}Y_{\uparrow\downarrow}Y^{\prime}_{\uparrow\downarrow} & :\,\text{chiral},
    \\
    X_{\uparrow}X^{\prime}_{\uparrow}-\delta_{\nu  x}Y_{\uparrow\uparrow}Y^{\prime}_{\uparrow\uparrow}     & :\,\text{helical},
  \end{dcases}
\end{align}
where we define $X_{s}=\ab(1/\sqrt{N_{\parallel}})\sum_{p}\mathrm{Im}\ab[g^{R}_{pss}\ab(E)]$ and $Y_{ss^{\prime}}=\ab(1/\sqrt{N_{\parallel}})\sum_{p}\mathrm{Re}\ab[f^{R}_{pss^{\prime}}\ab(E)]$, with primed quantities evaluated at $E+\hbar\omega$.
A detailed derivation is given in Appendix~\ref{pcal}.

\begin{figure}
  \begin{subfigure}{\columnwidth}
    \caption{Chiral}
    \includegraphics[width=0.8\columnwidth]{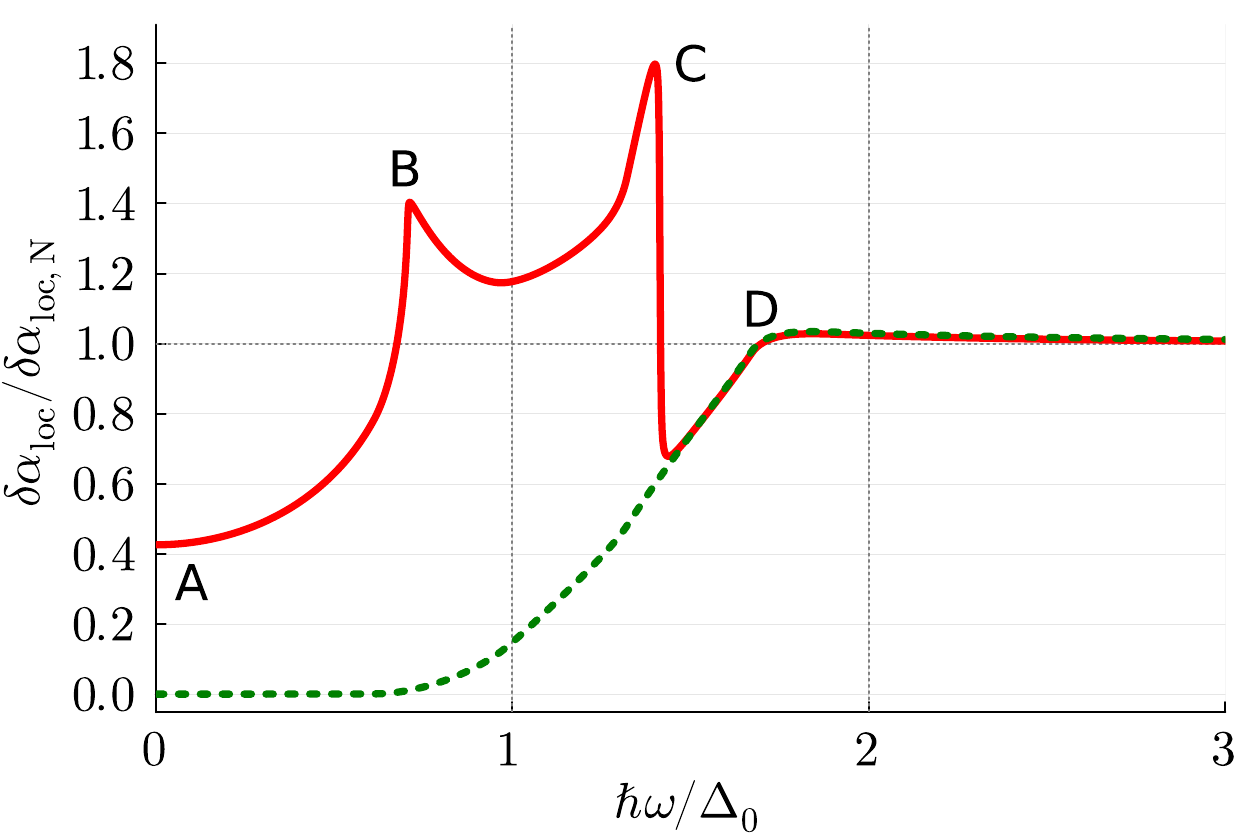}
    \label{cpGDC}
  \end{subfigure}

  \begin{subfigure}{\columnwidth}
    \caption{Helical}
    \includegraphics[width=0.8\columnwidth]{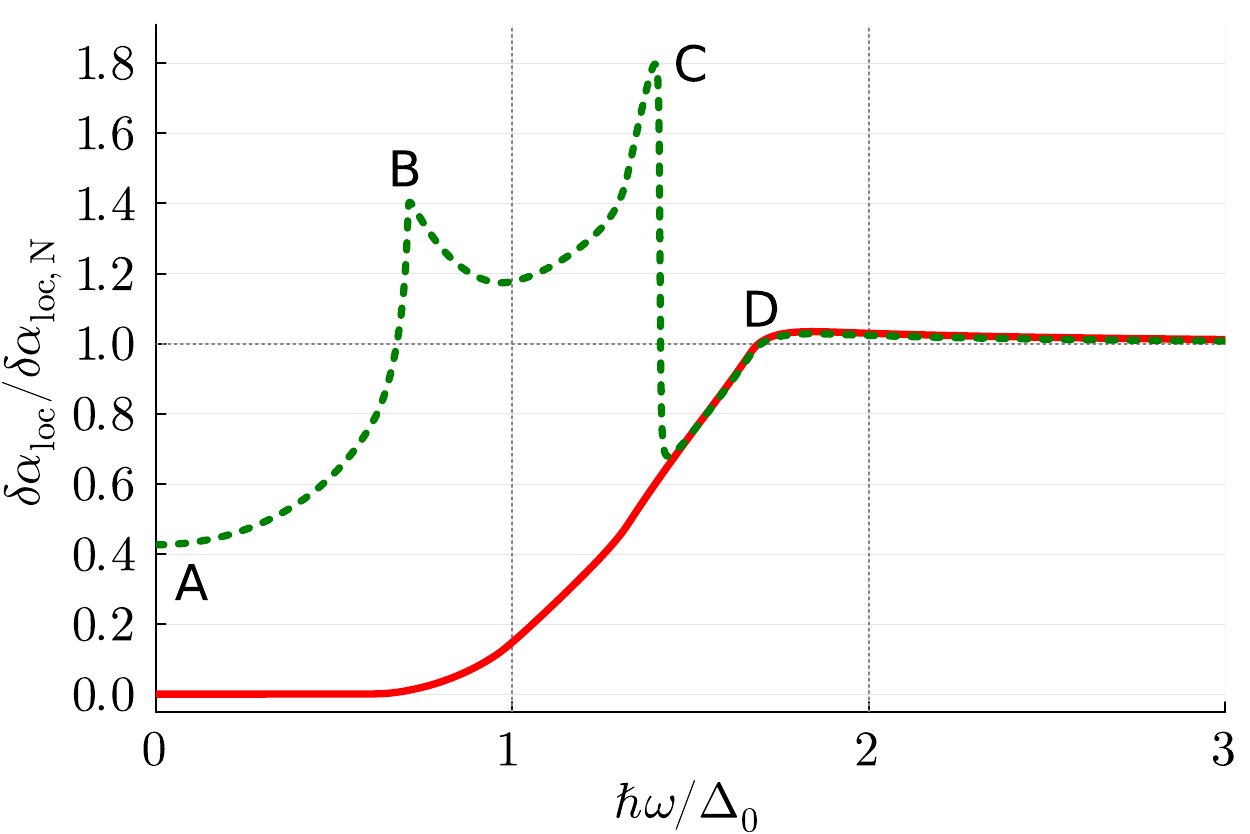}
    \label{hpGDC}
  \end{subfigure}
  
  \vspace{0.5em}

  \makebox[\linewidth]{%
    \includegraphics[trim=5cm 0.3cm 5cm 0.3cm,clip,scale=0.55]{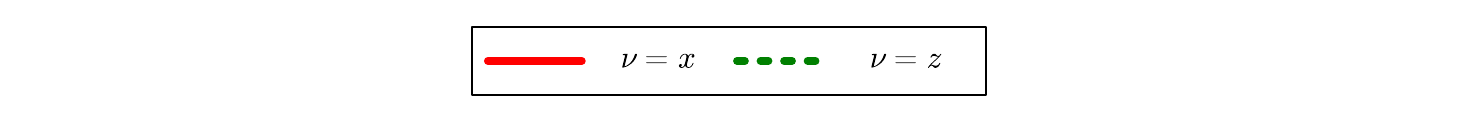}}

  \begin{subfigure}{\columnwidth}
    \caption{}
    \includegraphics[width=0.8\columnwidth]{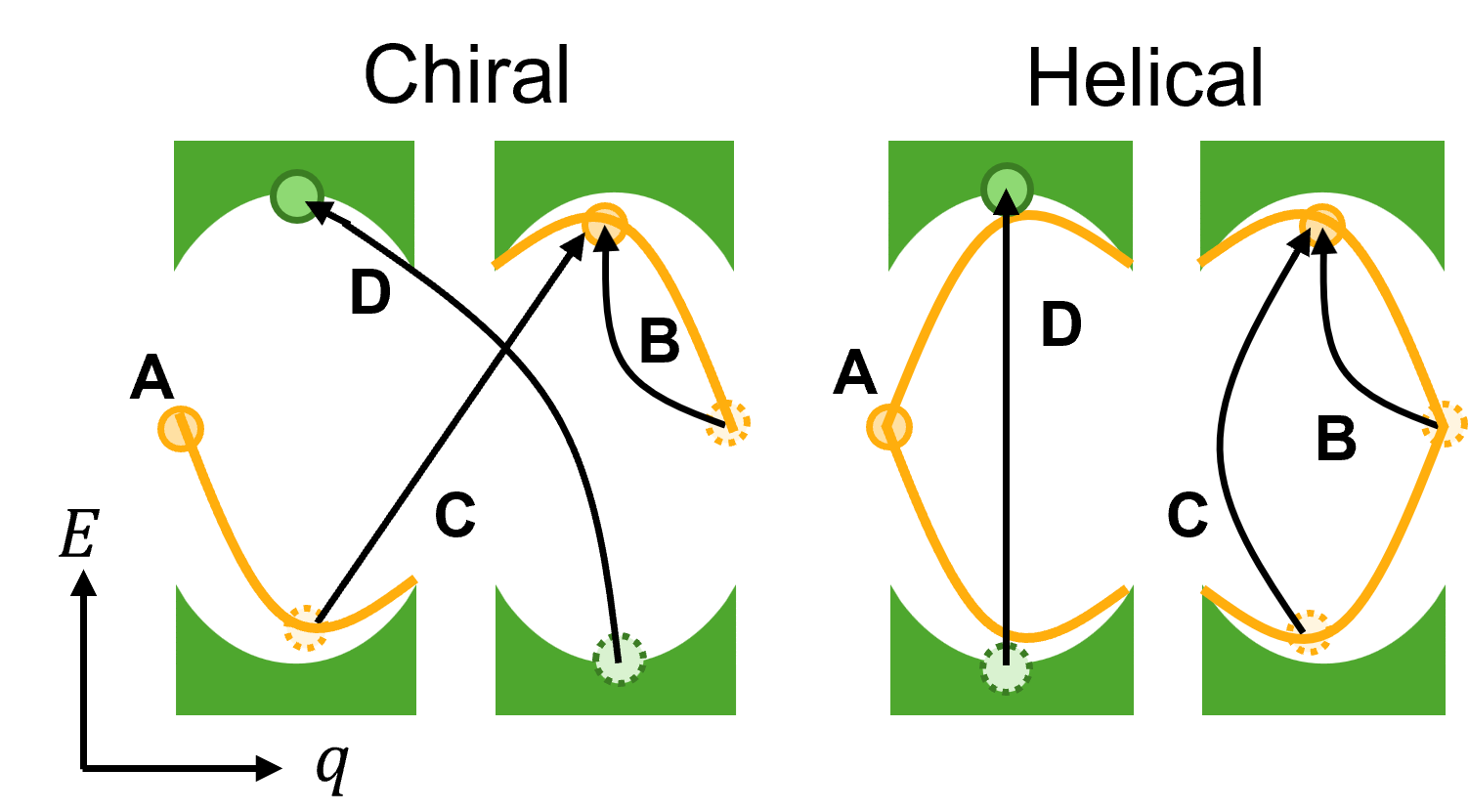}
    \label{p+ipexcitationloc}
  \end{subfigure}
  \caption{Frequency dependence of $\delta\alpha_{\mathrm{loc}}\ab(\omega)$ for $T=0$ along the $x$ and $z$ directions of the FI magnetization for the chiral-pSC and helical-pSC in Figs.~\ref{cpGDC} and \ref{hpGDC}. 
  In (a) the $\nu=y$ curve coincides with $\nu=x$ and in (b) the $\nu=y$ curve coincides with $\nu=z$. The labels ``A", ``B", ``C" and ``D" correspond to the schematic illustration of each excitation in Fig.~\ref{p+ipexcitationloc}.}
\end{figure}

The frequency dependence of $\delta\alpha_{\mathrm{loc}}$ at $T=0$ is shown in Figs.~\ref{cpGDC} and \ref{hpGDC} for the chiral-pSC and helical-pSC, respectively.
For the present model with identical parameters, the local damping contributions of the chiral and helical states coincide upon interchanging the FI magnetization directions $x\leftrightarrow z$.
The products of momentum-summed normal and anomalous spectral functions in Eq.~\eqref{eq:surface-chiloc} are identical for the two states, while the anomalous term contributes only for $\nu=z$ in the chiral state and $\nu=x$ in the helical state.
For the FI magnetization transverse to the respective edge-spin axes ($\nu=x,y$ for the chiral state and $\nu=y,z$ for the helical state), $\delta\alpha_{\mathrm{loc}}\ab(\omega)$ remains finite as $\hbar\omega\to0$ and exhibits two peaks (``B'', ``C'') and a shoulder (``D''), whereas for the magnetization along the axis ($\nu=z$ and $\nu=x$, respectively), it vanishes as $\hbar\omega\to0$ and exhibits only the shoulder (``D'').
The finite value in $\hbar\omega\to0$ indicates the excitation between ZEABS at the Brillouin zone boundary, denoted by ``A" in Fig.~\ref{p+ipexcitationloc}, and can be explained because the normal Green's function becomes a constant value by summing over the wave vector.
The two peaks emerging in $\hbar\omega/\Delta_{0}\sim0.7$ and $\hbar\omega/\Delta_{0}\sim1.4$ originate from edge-to-edge excitations, corresponding to ``B" and ``C" in Fig.~\ref{p+ipexcitationloc}, respectively.
The shoulder at $\hbar\omega/\Delta_{0}\sim1.8$ corresponds to the bulk-to-bulk excitation labeled ``D" in Fig.~\ref{p+ipexcitationloc}.

\begin{figure}
  \centering

  \begin{minipage}{0.48\linewidth}
    \subcaption{Chiral}
    \includegraphics[width=\columnwidth]{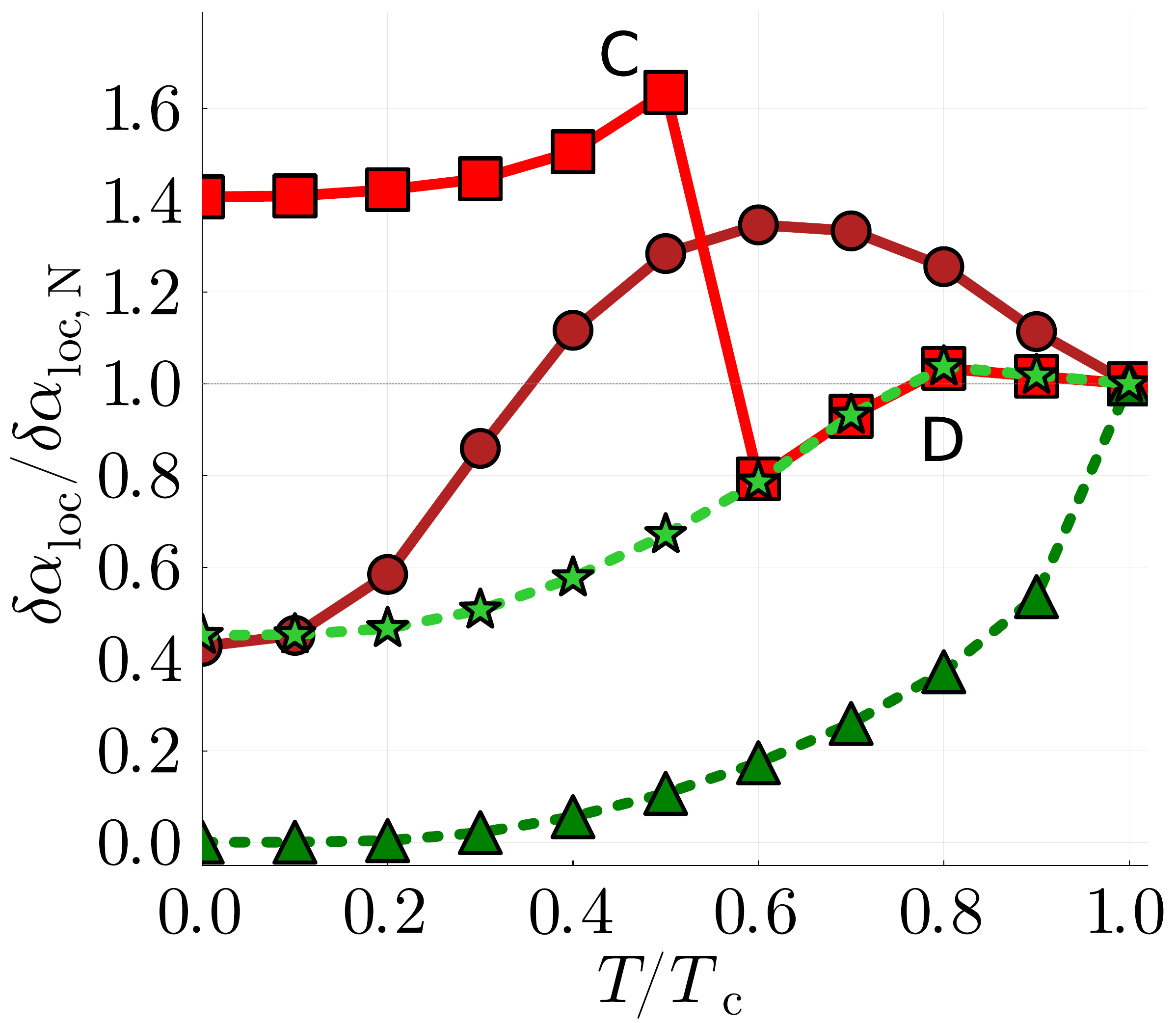}
    \label{cpGDCtemp}
  \end{minipage}
  \begin{minipage}{0.48\linewidth}
    \subcaption{Helical}
    \includegraphics[width=\linewidth]{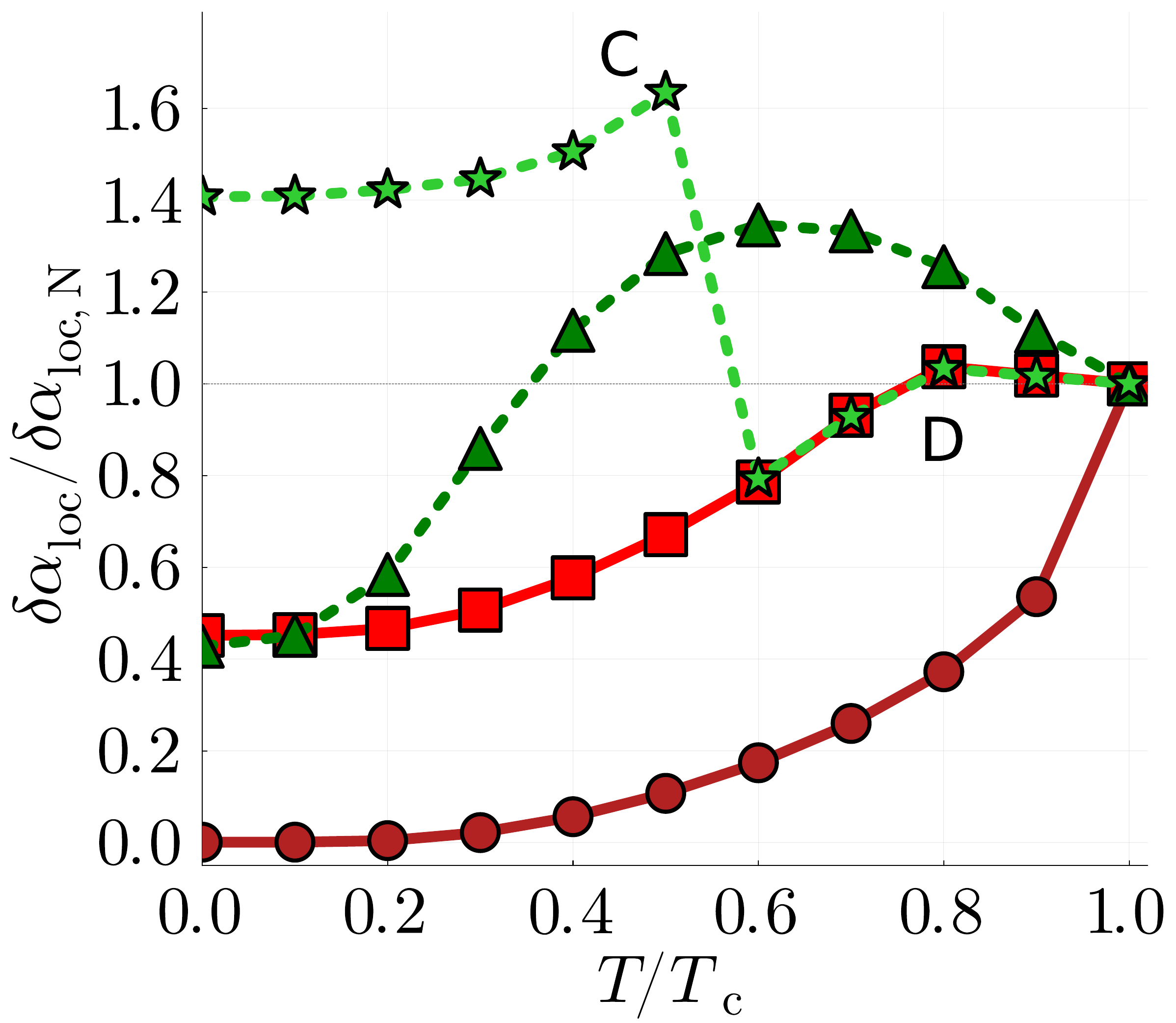}
    \label{hpGDCtemp}
  \end{minipage}
  \makebox[\linewidth]{%
    \includegraphics[trim=6cm 0.3cm 6cm 0.3cm,clip,scale=0.6]{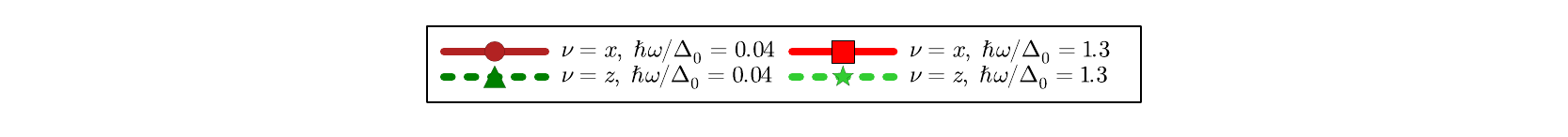}}
  \caption{Temperature dependence of $\delta\alpha_{\mathrm{loc}}\ab(\omega)$ along the $x$ and $z$ directions of FI magnetization for the chiral-pSC and helical-pSC in Figs.~\ref{cpGDCtemp} and \ref{hpGDCtemp}.}
  \label{GDCtemp}
\end{figure}

The temperature dependence of $\delta\alpha_{\mathrm{loc}}\ab(\omega)$ at low frequency $\ab(\hbar\omega/\Delta_{0}=0.04)$ and intermediate frequency $\ab(\hbar\omega/\Delta_{0}=1.3)$ is shown for chiral-pSC and helical-pSC in Figs.~\ref{cpGDCtemp} and \ref{hpGDCtemp}, respectively.
The two states exhibit the same behavior upon interchanging the FI magnetization directions $x\leftrightarrow   z$.
For the FI magnetization transverse to the edge-spin axes ($\nu=x,y$ for the chiral state and $\nu=y,z$ for the helical state), $\delta\alpha_{\mathrm{loc}}\ab(\omega)$ remains finite when $T=0$.
At low frequency, it exhibits a broad maximum as temperature increases, whereas at intermediate frequency, it exhibits two features corresponding to the excitations labeled ``C" and ``D" in Fig.~\ref{p+ipexcitationloc}.
For the FI magnetization along the edge-spin axis ($\nu=z$ and $\nu=x$, respectively), the edge-to-edge excitation is suppressed, whereas the edge-to-bulk and bulk-to-bulk excitations remain.

\section{Discussion}\label{discussion}

\subsection{Ising spin response of Majorana edge states}
To explain the low-energy suppression of edge-to-edge excitations, we consider a homogeneous edge of a two-dimensional superconductor with two Majorana fields~\cite{PhysRevB.82.180505,PhysRevB.110.174521}.
They satisfy $\gamma_a^\dagger=\gamma_a$ and $\{\gamma_a(x),\gamma_b(x')\}=\delta_{ab}\delta(x-x')$, where $a=1,2$ labels modes rather than physical spin.
With $\boldsymbol\gamma=(\gamma_1,\gamma_2)^T$, the effective edge Hamiltonians in magnetic fields are
\begin{align}
  H_{\mathrm e}^{\mathrm c}
   & =-\frac{i}{2}\int dx\,\boldsymbol\gamma^T
  \begin{pmatrix}
    \hbar v_{\mathrm c}\partial_x & g_z h_z^{\mathrm e}           \\
    -g_z h_z^{\mathrm e}          & \hbar v_{\mathrm c}\partial_x
  \end{pmatrix}
  \boldsymbol\gamma,
  \label{eq:edge-chiral}
  \\
  H_{\mathrm e}^{\mathrm h}
   & =-\frac{i}{2}\int dx\,\boldsymbol\gamma^T
  \begin{pmatrix}
    \hbar v_{\mathrm h}\partial_x & g_x h_x^{\mathrm e}            \\
    -g_x h_x^{\mathrm e}          & -\hbar v_{\mathrm h}\partial_x
  \end{pmatrix}
  \boldsymbol\gamma.
  \label{eq:edge-helical}
\end{align}
Here $x$ runs along the edge, $v_{\mathrm c,h}$ are edge velocities, and $g_i$ denote effective $g$ factors for the projected edge spin.
We absorb the Bohr magneton $\mu_B$ into $h_i^{\mathrm e}$, $h_i^{\mathrm e}=\mu_B B_i^{\mathrm e}$, where $B_i^{\mathrm e}$ are effective magnetic fields.
The field-independent chiral mixing is set to zero, as in the present spin-independent model.
Time reversal forbids a zero-field mass for the counterpropagating helical pair.
We retain first-derivative kinetic terms and zeroth-derivative spin couplings.

The chiral state with $\vb{d}\parallel\hat{\vb{z}}$ is even under $M_z=IC_{2z}$, whereas the helical state with $\vb{d}\perp\hat{\vb{z}}$ considered here belongs to the $A_{2u}$ representation of $D_{4h}$ and is even under $M_x=IC_{2x}$.
Here $I$ denotes inversion, and we assume that the normal-state Hamiltonian and boundary preserve the respective mirrors.
Each mirror therefore acts without an additional gauge transformation and satisfies $M_i^2=-1$ on the Majorana fields~\cite{PhysRevLett.111.087002}, allowing the representations
\begin{align}
  M_z:\ \boldsymbol\gamma(x) & \mapsto J\boldsymbol\gamma(x),
  \notag                                                       \\
  M_x:\ \boldsymbol\gamma(x) & \mapsto J\boldsymbol\gamma(-x),
  \qquad J=\begin{pmatrix}0&-1\\1&0\end{pmatrix}.
  \label{eq:edge-mirrors}
\end{align}
Equations~\eqref{eq:edge-chiral} and \eqref{eq:edge-helical} identify the spin density associated with $J$ as $-i\boldsymbol\gamma^T J\boldsymbol\gamma/2=i\gamma_1\gamma_2$.
This is the only nonconstant local bilinear without derivatives and is mirror even.
Since physical spin is axial, $M_z$ reverses $s_x,s_y$, whereas $M_x$ reverses $s_y,s_z$.
To this order, we normalize the effective edge spin densities as
\begin{equation}
  \vb{s}_{\mathrm e}^{\mathrm c}=i\gamma_1\gamma_2(0,0,1),\qquad
  \vb{s}_{\mathrm e}^{\mathrm h}=i\gamma_1\gamma_2(1,0,0).
  \label{eq:edge-ising-spin}
\end{equation}
These selection rules also apply with spin--orbit coupling when the full pairing matrix and uniform boundary preserve the specified mirrors and the two-mode subspace is symmetry closed.
For a two-dimensional chiral $p$-wave superfluid, Stone and Roy derived the coupling of a local bilinear of two Majorana edge fields to the component of a spin gauge field along the $d$-vector~\cite{PhysRevB.69.184511}.
For superconductors with time-reversal symmetry, mirror-symmetry constraints on the surface spin susceptibility have been related to odd-frequency pairing~\cite{PhysRevB.90.184506}, and magnetic responses of Majorana Kramers pairs have been classified using order-two symmetries~\cite{PhysRevB.103.094508}.
Crystal-symmetry selection rules for magnetic multipoles of helical Majorana surface states have also been formulated for three-dimensional time-reversal-invariant superconductors~\cite{PhysRevLett.123.097002}.
This classification has been extended using magnetic point groups to pairs of degenerate Majorana modes at high-symmetry points in the surface Brillouin zone of three-dimensional superconductors with broken time-reversal symmetry~\cite{PhysRevB.110.134518}.
An analogous Ising-like edge spin polarized along the $d$-vector was predicted in the low-temperature regime of chiral $p$-wave superfluid ${}^{3}$He-A confined in a slab~\cite{JPSJ.79.113601}.

Since FMR probes spin transverse to the FI magnetization [Eq.~\eqref{sigmapm}], these selection rules imply that the leading edge-to-edge response vanishes for $\nu=z$ (chiral) and $\nu=x$ (helical), where the FI magnetization is parallel to both the interface plane and the $d$-vector for the zero-energy states.
For the present model, the same suppression is also obtained from the edge contributions to the surface Green's functions in Appendix~\ref{app:edge-gf}.
Derivative corrections and processes involving bulk states are outside this leading projection.

\subsection{Identification of Cooper-pair symmetry from uniform and local damping contributions}
\begin{table}
  \caption{Signatures distinguishing the chiral and helical states on the $(010)$ surface. The $\delta\alpha_{\mathrm{uni}}$ rows give the edge-to-edge contribution for the FI magnetization transverse to the edge-spin axis within the linear-edge regime: the frequency dependence at $T=0$ for $2\delta\ll\hbar\omega$, and the temperature dependence at fixed low frequency for $\hbar\omega\ll4k_{\mathrm{B}}T$ (for $4k_{\mathrm{B}}T\ll\hbar\omega$ the helical contribution also saturates). The $\delta\alpha_{\mathrm{loc}}$ row gives the $\hbar\omega\to0$ limit at $T=0$ for the same magnetization directions. The last row gives the FI magnetization direction for which the low-energy edge-to-edge contribution is suppressed, i.e., the edge-state spin axis, which is parallel to the $d$-vector for the zero-energy states and the interface.
  }
  \begin{ruledtabular}
  \begin{tabular}{lcc}
     & Chiral & Helical \\
    \hline
    $\delta\alpha_{\mathrm{uni}}\ab(\omega,\,T=0;\,2\delta\ll\hbar\omega)$ & $\propto\,\delta/(\hbar\omega)^{2}$ & $\propto\,1/\hbar\omega$ \\
    $\delta\alpha_{\mathrm{uni}}\ab(\omega,\,T;\,2\delta\ll\hbar\omega\ll4k_{\mathrm{B}}T)$ & $\propto\,T^{0}$ & $\propto\,1/T$ \\
    $\delta\alpha_{\mathrm{loc}}\ab(\omega,\,T=0;\,\hbar\omega\to0)$ & finite & finite \\
  Edge response suppressed for & $\nu=z$ & $\nu=x$ 
  \end{tabular}
  \end{ruledtabular}
  \label{table}
\end{table}

\paragraph{FMR-based identification of Cooper-pair symmetry}
We discuss how FMR modulation can be used to identify the Cooper-pair symmetry, as summarized in Table~\ref{table}.
The total damping enhancement consists of  the uniform and local contributions, $\delta\alpha_{\mathrm{uni}}\ab(\omega)$ and $\delta\alpha_{\mathrm{loc}}\ab(\omega)$, respectively.
For an ideal disorder-free interface, $J_{1}\gg J_{2}$, the uniform contribution $\delta\alpha_{\mathrm{uni}}\ab(\omega)$ dominates the FMR response.
In this case, the chiral and helical states can be distinguished by their distinct frequency and temperature dependences, as well as their spin anisotropy.
Increasing \(J_2\) at fixed \(J_1\) enhances the relative weight of the local contribution $\delta\alpha_{\mathrm{loc}}\ab(\omega)$ to the total damping.
In this regime, some of the frequency- and temperature-dependent responses of the chiral and helical states become similar.
Nevertheless, the characteristic frequency dependence of the edge-to-edge contribution remains a useful signature of the pairing symmetry, because the local contribution $\delta\alpha_{\mathrm{loc}}\ab(\omega)$ only adds a frequency-independent background at low frequency [Figs.~\ref{cpGDC} and \ref{hpGDC}].
For the FI magnetization transverse to the edge-spin axis (e.g., $\nu=y$ for both states), in the linear-edge regime satisfying $2\delta\ll\hbar\omega$ and $4k_{\mathrm{B}}T\ll\hbar\omega$, an $\ab(\hbar\omega)^{-2}$ increase of $\delta\alpha$ toward low frequency indicates the chiral state, whereas an $\ab(\hbar\omega)^{-1}$ increase indicates the helical state [Eqs.~\eqref{eq:uniform-chiral-edge} and \eqref{eq:uniform-helical-edge}].

\paragraph{Distinct signatures of chiral and helical states}
The two signatures are not equivalent in magnitude: the helical $\ab(\hbar\omega)^{-1}$ term is a property of the edge spectrum and is independent of $\delta$, whereas the chiral $\ab(\hbar\omega)^{-2}$ term originates from the quasiparticle broadening and is smaller by a factor of order $\delta/\hbar\omega$, so that the $\ab(\hbar\omega)^{-1}$ term is the more readily observable of the two.
The low-frequency temperature dependence ($\hbar\omega\ll4k_{\mathrm{B}}T$) provides a complementary signature. For the FI magnetization transverse to the edge-spin axis, the chiral response saturates upon cooling, whereas the helical response increases as $1/T$ [Eq.~\eqref{eq:uniform-helical-edge}] down to $T\sim\hbar\omega/4k_{\mathrm{B}}$, below which it also saturates. Because the local contribution has the same temperature dependence for both states and saturates upon cooling (Fig.~\ref{GDCtemp}), the $1/T$ increase of the helical state remains visible as long as the uniform contribution is not negligible.
The frequency and temperature dependences of the helical edge contribution are two limits of a single scaling form, $\hbar\omega\delta\alpha^{\mathrm{h}}_{\mathrm{uni,edge}}\propto\tanh(\hbar\omega/4k_{\mathrm{B}}T)$ [Eq.~\eqref{eq:uniform-helical-edge}], a function of $\hbar\omega/k_{\mathrm{B}}T$ alone, whereas the chiral edge contribution is independent of temperature.
The spin anisotropy provides another robust signature because the local contribution exhibits the same characteristic anisotropy as the uniform contribution.

\paragraph{Comparison with previous work}
This robustness of the spin anisotropy contrasts with the bulk analysis of Ref.~\cite{PhysRevB.106.L161406}, in which interfacial disorder renders the bulk response isotropic.
Thus, the edge states provide additional information for identifying the Cooper-pair symmetry.

\subsection{Experimental considerations}
We finally note the experimental possibility with FMR modulation in the FI/$p$SC bilayer system.
FMR modulation can measure $\mathrm{GHz}$ range, which corresponds to the energy scale: $20\,\mathrm{GHz}\approx1\,\mathrm{K}$.
For a pSC with a transition temperature \(T_{\mathrm{c}}\sim1\,\mathrm{K}\), the excitation peaks considered here correspond to frequencies in the microwave range.
However, as the microwave frequency $\omega$ is increased, the corresponding resonance field $\omega_{\vb{k}=\vb{0}}$ also increases, requiring a larger static magnetic field $h_{\mathrm{dc}}$ to be applied to the FI because $\omega_{\vb{k}=\vb{0}}=\gamma    h_{\mathrm{dc}}$.
When the applied field becomes comparable to the upper critical magnetic field, $h_{\mathrm{dc}}\sim  H_{\mathrm{c2}}$, the contribution of the vortex states can no longer be neglected, discussed in Ref.~\cite{PhysRevB.106.L161406}.
Consequently, the characteristic response in high frequency range becomes obscured by the vortex contribution.
Accordingly, it is important to focus on the  distinctive frequency scalings of the edge-to-edge excitations in the low-frequency regime and the spin anisotropy depending on the magnetization directions of the FI.
\section{Conclusion}\label{Conclusion}
In this paper, we have theoretically investigated FMR modulation of chiral-pSC and helical-pSC by evaluating the enhancement of the Gilbert damping constant $\delta\alpha$ arising from edge and bulk quasiparticle excitations.

In the flat-interface limit of momentum-conserving exchange, the low-energy edge response at $T=0$ exhibits distinct frequency dependences for FI magnetization transverse to the edge-spin axis.
Within the linear-edge regime and for $2\delta\ll\hbar\omega$, the chiral and helical edge contributions scale as $\delta/(\hbar\omega)^2$ and $1/(\hbar\omega)$, respectively.
Within the linear-edge approximation with fixed velocities, spectral weights, and broadening, the occupation factors leave the chiral edge contribution temperature independent.
The helical edge contribution saturates as $T\to0$ at fixed frequency and exhibits a $1/T$ dependence for $2\delta\ll\hbar\omega\ll 4k_{\mathrm{B}}T$.
These frequency and temperature dependences provide signatures for distinguishing the two pairing states when the uniform edge contribution can be resolved.

On the $(010)$ surface, the leading low-energy edge response is suppressed when the FI magnetization is aligned with the edge-spin axis, which is $z$ for the chiral state and $x$ for the helical state.
With a clean superconducting boundary, this Ising-like anisotropy survives both momentum-conserving and momentum-randomizing interfacial exchange.
These results provide complementary frequency and angular signatures for distinguishing the two pairing states considered here and constraining their $d$-vector structures.

Future FMR experiments on candidate spin-triplet superconductors could test these signatures and complement existing probes of unconventional pairing by measuring the spin dynamics of superconducting edges.

\acknowledgments
S.M. is supported by JST SPRING (Grant No.~JPMJSP2125).
Y.O. is supported by JSPS KAKENHI (Grant No.~JP26K06985).
T.K. is supported by JSPS KAKENHI (Grant No.~JP24K06951).
M.M. is supported by the National Natural Science Foundation of China (NSFC) (Grant No.~12374126), the Priority Program of the Chinese Academy of Sciences (Grant No.~XDB28000000), and JSPS KAKENHI (Grant Nos.~JP23H01839, JP24H00322, and JP26H02144).
A.Y. is supported by JSPS KAKENHI (Grant Nos.~JP24H00853 and JP25K07224).

\appendix

\section{Derivation of surface Green's function}\label{SGF}
The surface Gor'kov--Green function in the semi-infinite system written in Eq.~\eqref{GGF} obeys the following equation \cite{umerski1997closed}:
\begin{align}
    \tilde{g}^{R}_{p}\ab(E)
     & =
    \ab(E+i\delta-\vb{u}_{p}-\vb{t}^{\dag}\tilde{g}^{R}_{p}\ab(E)\vb{t})^{-1}
    \notag
    \\
     & \equiv
    X^{\infty}_{\bullet}\vb{0},
\end{align}
where $X$ is the $2N\times2N$ matrix given by
\begin{align}
    X
    =
    \begin{pmatrix}
        0              & \vb{t}^{-1}
        \\
        -\vb{t}^{\dag} & (E+i\delta-\vb{u}_{p})\vb{t}^{-1}
    \end{pmatrix},
\end{align}
and ``$\bullet$" the operator of the M\"obius transformation defined as
\begin{align}
    \begin{pmatrix}
        a & b
        \\
        c & d
    \end{pmatrix}
    _{\bullet}
    Y
     & =
    (aY+b)(cY+d)^{-1}.
\end{align}
The surface Green's function can be derived as
\begin{equation}
    \tilde{g}^{R}_{p}\ab(E)=\vb{O}_{12}\vb{O}^{-1}_{22},
\end{equation}
from the matrix $\vb{O}$ written as
\begin{equation}
    \vb{O}
    =
    \begin{pmatrix}
        \vb{O}_{11} & \vb{O}_{12}
        \\
        \vb{O}_{21} & \vb{O}_{22}
    \end{pmatrix},
\end{equation}
which satisfies $\vb{O}^{-1}X\vb{O}=\mathrm{diag}\ab(\lambda_{1},\lambda_{2},\cdots,\lambda_{2N})$ and $|\lambda_{1}|<|\lambda_{2}|<\cdots<|\lambda_{2N}|$.

\section{Normal and anomalous edge Green's functions}\label{app:edge-gf}
A low-energy relation between normal and anomalous Green's functions arising from a Majorana zero mode has been derived for a one-dimensional $p_x$-wave superconductor~\cite{PhysRevB.87.104513}.
For the spin-independent lattice model in Eq.~\eqref{pSCHamiltonian}, the relation between the normal and anomalous edge Green's functions follows from the subspace spanned by zero modes of $H_0(p)$ localized at the edge.
At fixed surface momentum $p$, write the semi-infinite BdG operator as $H_0(p)+V_p$, where $H_0$ has onsite term $\xi_p\tau^z$ and hopping $\vb{t}$, with
\begin{align}
    \vb{t} & =-t\tau^z+i\Delta_{\perp}\tau^xN,
           & \Delta_{\perp}                    & =\frac{\Delta}{2\sqrt{2}},
    \notag                                                                         \\
    V_p    & =-D_p\tau^yM,
           & D_p                               & =\frac{\Delta}{\sqrt{2}}\sin(pa).
\end{align}
Here $\tau^i$ act in Nambu space, and $(M,N)=(\sigma^x,\sigma^x)$ for the chiral state and $(\sigma^{0},-\sigma^z)$ for the helical state.
The operator
\begin{equation}
    \Gamma=\tau^yN,\qquad
    \Gamma^2=1,\qquad
    \{H_0,\Gamma\}=0
\end{equation}
allows the zero modes of $H_0(p)$, localized at the edge, to be chosen with $\Gamma\psi=\lambda\psi$.
For layers $n=0,-1,\ldots$, substituting $\psi_n\propto r^{-n}$ into $H_0(p)\psi=0$ gives
\begin{equation}\label{eq:edge-decay}
    (t-\lambda\Delta_{\perp})r^2-\xi_p r
    +(t+\lambda\Delta_{\perp})=0.
\end{equation}
For $t,\Delta_{\perp}>0$ and $|\xi_p|<2t$, the $\lambda=-1$ sector has two roots inside the unit circle, counting multiplicity.
Their decaying combination, or its repeated-root limit, satisfies the open-boundary condition $\psi_1=0$; the opposite sector has no such boundary zero mode.

Since $[M,N]=0$, the tangential term satisfies
\begin{equation}
    \{H_0,V_p\}=0,\qquad [\Gamma,V_p]=0.
\end{equation}
Thus $V_p$ preserves both $\ker H_0$ and its selected $\Gamma=-1$ sector.
Restoring $V_p$ therefore produces dispersing edge states in the same sector; $\Gamma$ does not anticommute with the full Hamiltonian at generic $p$.
Writing the surface components of a normalized edge eigenstate as $(u_{p\alpha},v_{p\alpha})^T$, we obtain
\begin{equation}\label{eq:edge-uv}
    v_{p\alpha}=-iN u_{p\alpha}.
\end{equation}
The edge-only spectral contributions consequently satisfy
\begin{align}
    g_{e,p}^R(E) & =\sum_{\alpha\in e}
    \frac{u_{p\alpha}u_{p\alpha}^{\dag}}
    {E-E_{p\alpha}+i0^+},
    \notag                           \\
    f_{e,p}^R(E) & =\sum_{\alpha\in e}
    \frac{u_{p\alpha}v_{p\alpha}^{\dag}}
    {E-E_{p\alpha}+i0^+}
    =i g_e^R(p,E)N.
    \label{eq:edge-gf}
\end{align}
For the chiral state this gives
\begin{equation}
    \operatorname{Re}f^R_{e,p\uparrow\downarrow}(E)
    =-\operatorname{Im}g^R_{e,p\uparrow\uparrow}(E),
\end{equation}
whereas for the helical state,
\begin{align}
     & \operatorname{Re}\!\left[
                              f^R_{e,p\uparrow\uparrow}(E)-f^R_{e,p\downarrow\downarrow}(E)\right]
    \notag                                                                                                  \\
     & \qquad=\operatorname{Im}\!\left[
                                      g^R_{e,p\uparrow\uparrow}(E)+g^R_{e,p\downarrow\downarrow}(E)\right].
\end{align}
Substitution at both energies and momenta into the kernels $\mathcal{K}^{\mathrm{c}}_{z}$ and $\mathcal{K}^{\mathrm{h}}_{x}$ in Eq.~\eqref{eq:surface-chi-kernels} cancels the normal and anomalous edge-to-edge terms for $\nu=z$ in the chiral state and $\nu=x$ in the helical state, respectively.
Equation~\eqref{eq:edge-gf} uses the pairing phase and termination of this lattice model; the opposite surface reverses the selected $\lambda$ and the corresponding phase relation.
It is not a general consequence of particle-hole symmetry in the presence of spin--orbit coupling.
The full surface Green's functions also contain bulk contributions and do not satisfy this edge-only identity.

\section{Derivation of the surface spin susceptibility}\label{pcal}
We derive Eqs.~\eqref{eq:surface-chi-common} and \eqref{eq:surface-chi-kernels} for the fixed-order-parameter BdG Hamiltonian in Eq.~\eqref{pSCHamiltonian}.
The surface Green's functions used below include both edge and bulk contributions.
Throughout this appendix, $\sigma^\pm$ denote the ladder matrices in Eq.~\eqref{sigmapm} for the chosen magnetization direction $\nu$.

\subsection{Matsubara response and spectral representation}
Let $\epsilon_n=(2n+1)\pi/\beta$ and $\Omega_m=2m\pi/\beta$ denote fermionic and bosonic Matsubara energies, respectively.
Wick's theorem gives the connected spin correlation function as
\begin{align}
    \chi_{q,\nu}(i\Omega_m)
     & =-\frac{1}{4\beta}\sum_{p,n}
    \operatorname{tr}\Big[\sigma^+ g'\sigma^- g
                         -\sigma^+ f'(\sigma^-)^T\bar f\Big],
    \label{eq:app-chi-matsubara}
\end{align}
where the trace is over spin.
Here $g=g_p(i\epsilon_n)$ and $g'=g_{p+q}(i\epsilon_n+i\Omega_m)$, with the same convention for the anomalous functions.
The relative minus sign in the brackets follows from the anomalous Wick contraction.
Since Eq.~\eqref{eq:app-chi-matsubara} is expressed directly in terms of electron operators, no additional factor of $1/2$ from Nambu doubling is required.

For either superconducting state, the real-space BdG matrix at fixed $p$ obeys $\tau^z H_p^*\tau^z=H_p$ in the pairing-phase convention of Eq.~\eqref{pSCHamiltonian}.
Together with Hermiticity, this yields the following relations for the surface Nambu Green's function $\tilde g$ defined in Eq.~\eqref{GGF}, with the common real-energy argument $E$ suppressed:
\begin{align}
    \tilde g_p^A                    & =[\tilde g_p^R]^\dagger,
                                     &
    \tau^z[\tilde g_p^R]^*\tau^z    & =\tilde g_p^A,
    \notag                                                        \\
    f_p^A                           & =-[f_p^R]^*,
                                     &
    \bar f_p^R                      & =-[f_p^R]^T.
    \label{eq:app-gf-symmetry}
\end{align}
The Pauli matrix $\tau^z$ acts in Nambu space.
Define the real spin matrices $A=\operatorname{Im}g_p^R(E)$ and $B=\operatorname{Re}f_p^R(E)$.
For $\rho_X=-(X^R-X^A)/(2\pi i)$, Eq.~\eqref{eq:app-gf-symmetry} gives
\begin{equation}
    \rho_g=-\frac{A}{\pi},\qquad
    \rho_f=\frac{iB}{\pi},\qquad
    \rho_{\bar f}=-\frac{iB^T}{\pi}.
    \label{eq:app-spectral-blocks}
\end{equation}
Thus the anomalous spectral weight involves $\operatorname{Re}f^R$, rather than $\operatorname{Im}f^R$.
This form depends on the specified pairing phase.

Using the spectral representation of each Green's function and
\begin{align}
     & \frac{1}{\beta}\sum_n
    \frac{1}{i\epsilon_n-E}
    \frac{1}{i\epsilon_n+i\Omega_m-E'}
    =\frac{n_{\mathrm F}(E)-n_{\mathrm F}(E')}
    {i\Omega_m+E-E'},
\end{align}
we analytically continue $i\Omega_m\to\Omega+i0^+$, where $\Omega=\hbar\omega$.
Taking the imaginary part with
$\operatorname{Im}[-(x+i0^+)^{-1}]=\pi\delta(x)$ then gives
\begin{align}
    \operatorname{Im}\chi^R_{q,\nu}(\omega)
     & =\sum_p\int\frac{dE}{\pi}
                  [n_{\mathrm F}(E)-n_{\mathrm F}(E+\Omega)]
    \notag                                                   \\  & \qquad\times
    \mathcal K_\nu(p,E;q,\omega),
    \label{eq:app-chi-spectral}                              \\
    \mathcal K_\nu
     & =\frac{1}{4}\ab(\operatorname{tr}[\sigma^+ A'\sigma^- A]
    -\operatorname{tr}[\sigma^+ B'(\sigma^-)^T B^T]),
    \label{eq:app-chi-trace}
\end{align}
where the primed matrices are evaluated at $(p+q,E+\Omega)$.

\subsection{Spin structure of the chiral and helical kernels}
For the chiral state, $H_p$ commutes with $\tau^z\sigma^z$ and $\sigma^x$, where identities in the other spaces are implicit.
These symmetries give
\begin{equation}
    g_p^R=g_{p\uparrow\uparrow}^R\sigma^0,\qquad
    f_p^R=f_{p\uparrow\downarrow}^R\sigma^x.
\end{equation}
Hence $A=A_\uparrow\sigma^0$ and $B=B_{\uparrow\downarrow}\sigma^x$.
The spin traces are
\begin{equation}
    \operatorname{tr}[\sigma^+\sigma^-]=4,\qquad
    \operatorname{tr}[\sigma^+\sigma^x(\sigma^-)^T\sigma^x]
    =4\delta_{\nu z}.
\end{equation}
Equation~\eqref{eq:app-chi-trace} reduces to
\begin{equation}
    \mathcal K_\nu^{\mathrm c}
    =A_\uparrow A'_\uparrow
    -\delta_{\nu z}B_{\uparrow\downarrow}B'_{\uparrow\downarrow}.
\end{equation}

For the helical state, $[H_p,\sigma^z]=0$ makes both $g_p^R$ and $f_p^R$ diagonal in spin.
Thus $A=\operatorname{diag}(A_\uparrow,A_\downarrow)$ and $B=\operatorname{diag}(B_{\uparrow\uparrow},B_{\downarrow\downarrow})$.
The first trace in Eq.~\eqref{eq:app-chi-trace} evaluates to
\begin{align}
    \mathcal N_\nu
     & \equiv\frac{1}{4}\operatorname{tr}[\sigma^+ A'\sigma^- A]
    =\sum_{ab}\frac{|(\sigma^+)_{ab}|^2}{4} A_a A_b'
    \notag                                            \\
     & =
    \begin{cases}
        \dfrac{(A_\uparrow+A_\downarrow)
            (A'_\uparrow+A'_\downarrow)}{4}, & \nu=x,y, \\[1ex]
        A_\uparrow A'_\downarrow,            & \nu=z.
    \end{cases}
\end{align}
The second trace evaluates to
\begin{align}
    \mathcal F_\nu
     & \equiv\frac{1}{4}\operatorname{tr}[\sigma^+ B'(\sigma^-)^T B^T]
    \notag                                                \\
     & =
    \begin{cases}
        \dfrac{(B_{\uparrow\uparrow}-B_{\downarrow\downarrow})
            (B'_{\uparrow\uparrow}-B'_{\downarrow\downarrow})}{4},
           & \nu=x, \\[1ex]
        \dfrac{(B_{\uparrow\uparrow}+B_{\downarrow\downarrow})
            (B'_{\uparrow\uparrow}+B'_{\downarrow\downarrow})}{4},
           & \nu=y, \\[1ex]
        0, & \nu=z.
    \end{cases}
\end{align}
Taking $\mathcal K_\nu^{\mathrm h}=\mathcal N_\nu-\mathcal F_\nu$ reproduces the helical kernel in Eq.~\eqref{eq:surface-chi-kernels}.

\subsection{Local spin susceptibility}
The local response is $\chi_{\mathrm{loc},\nu}=N_\parallel^{-1}\sum_q\chi^R_{q,\nu}$.
For sums over the full Brillouin zone, the change of variables $k=p+q$ gives
\begin{align}
     & \frac{1}{N_\parallel}\sum_{p,q}U_p(E)V_{p+q}(E+\Omega)
    \notag                                                    \\
     & \quad=
    \left[\frac{1}{\sqrt{N_\parallel}}\sum_p U_p(E)\right]
    \left[\frac{1}{\sqrt{N_\parallel}}\sum_k V_k(E+\Omega)\right].
    \label{eq:app-local-factorization}
\end{align}
Consequently, the local kernels follow by replacing $A_s$ and $B_{ss'}$ with the quantities $X_s$ and $Y_{ss'}$ defined below Eq.~\eqref{eq:surface-chiloc}.
The factor $1/\sqrt{N_\parallel}$ in each quantity accounts for the normalization of the $q$ sum.

For the helical state, time-reversal symmetry together with Eq.~\eqref{eq:app-gf-symmetry} implies
\begin{equation}
    g^R_{p\uparrow\uparrow}=g^R_{-p\downarrow\downarrow},
    \qquad
    f^R_{p\uparrow\uparrow}=-f^R_{-p\downarrow\downarrow}.
\end{equation}
Summing over $p$ therefore gives $X_\uparrow=X_\downarrow$ and $Y_{\uparrow\uparrow}=-Y_{\downarrow\downarrow}$ at each energy.
The resulting local kernels are
\begin{align}
    \mathcal K^{\mathrm c}_{\mathrm{loc},\nu}
     & =X_\uparrow X'_\uparrow
    -\delta_{\nu z}Y_{\uparrow\downarrow}Y'_{\uparrow\downarrow},
    \notag                     \\
    \mathcal K^{\mathrm h}_{\mathrm{loc},\nu}
     & =X_\uparrow X'_\uparrow
    -\delta_{\nu x}Y_{\uparrow\uparrow}Y'_{\uparrow\uparrow},
\end{align}
with primed quantities evaluated at $E+\Omega$.
Substitution into the energy integral in Eq.~\eqref{eq:app-chi-spectral}, after the momentum sums have been performed, yields Eq.~\eqref{eq:surface-chiloc}.
The cancellation of the helical anomalous term for $\nu=y$ applies to the local response; it does not generally hold at fixed $q$.

\bibliography{ref}

\end{document}